\documentclass[iop]{emulateapj}
\usepackage{graphicx}
\usepackage{booktabs}
\usepackage{amsmath}
\usepackage{amssymb}
\usepackage{amsthm}
\usepackage{hyperref}
\usepackage[caption=false]{subfig}
\usepackage{color}

\usepackage{listings}
\newcommand{\nver}{\hat{\mathbf{n}}}
\newcommand{\cov}{\text{Cov}}
\newcommand{\Nsim}{N_{\text{sim}}}

\def\lsim{\,\lower2truept\hbox{${<\atop\hbox{\raise4truept\hbox{$\sim$}}}$}\,}
\def\gsim{\,\lower2truept\hbox{${> \atop\hbox{\raise4truept\hbox{$\sim$}}}$}\,}

\begin{document}
\title{Cross-correlation between the CMB lensing potential measured by \emph{Planck}\footnote{Based on observations obtained with \emph{Planck} (http://www.esa.int/Planck), an ESA science mission with instruments and contributions directly funded by ESA Member States, NASA, and Canada.} and high-$z$ sub-mm galaxies detected by the \emph{Herschel}-ATLAS survey\footnote{\textit{Herschel} is an ESA space observatory with science instruments provided by European-led Principal Investigator consortia and with important participation from NASA.}}


\author{F. Bianchini\altaffilmark{1,12}, P. Bielewicz\altaffilmark{1}, A. Lapi\altaffilmark{2,1,12,13}, J. Gonzalez-Nuevo\altaffilmark{3}, C. Baccigalupi\altaffilmark{1,12}, G. de Zotti\altaffilmark{4,1}, L. Danese\altaffilmark{1},
N. Bourne\altaffilmark{5}, A. Cooray\altaffilmark{6}, L. Dunne\altaffilmark{7,5}, S. Dye\altaffilmark{8}, S. Eales\altaffilmark{9}, R. Ivison\altaffilmark{5,10}, S. Maddox\altaffilmark{7,5}, M. Negrello\altaffilmark{4},  D. Scott\altaffilmark{11}, M. W. L. Smith\altaffilmark{9}, E. Valiante\altaffilmark{9}}

\altaffiltext{1}{Astrophysics Sector, SISSA, Via Bonomea 265, I-34136 Trieste, Italy; fbianchini@sissa.it}
\altaffiltext{2}{Dipartimento di Fisica, Universit\`a ``Tor Vergata'', Via della Ricerca Scientifica 1, I-00133 Roma, Italy}
\altaffiltext{3}{Inst. de Fisica de Cantabria (CSIC-UC), Avda. los Castros s/n, 39005 Santander, Spain}
\altaffiltext{4}{INAF - Osservatorio Astronomico di Padova, Vicolo dell'Osservatorio 5, I-35122 Padova, Italy}
\altaffiltext{5}{Institute for Astronomy, University of Edinburgh, Royal Observatory, Blackford Hill, Edinburgh EH9 3HJ, UK}
\altaffiltext{6}{Department of Physics and Astronomy, University of California Irvine CA 92697 USA}
\altaffiltext{7}{Department of Physics and Astronomy, University of Canterbury, Private Bag 4800, Christchurch, 8140, New Zealand}
\altaffiltext{8}{School of Physics and Astronomy, University of Nottingham, University Park, Nottingham, NG7 2RD, UK}
\altaffiltext{9}{School of Physics and Astronomy, Cardiff University, Queens Buildings, The Parade, Cardiff CF24 3AA, UK}
\altaffiltext{10}{European Southern Observatory, Karl Schwarzschild Strasse 2, Garching, Germany}
\altaffiltext{11}{Department of Physics \& Astronomy, University of British Columbia,Vancouver, BC V6T 1Z1, Canada}
\altaffiltext{12}{INFN - Sezione di Trieste, Via Valerio 2, I-34127 Trieste, Italy}
\altaffiltext{13}{INAF - Osservatorio Astronomico di Trieste, via Tiepolo 11, 34131, Trieste, Italy}

\email{fbianchini@sissa.it}


\begin{abstract}

We present the first measurement of the correlation between the map of the cosmic microwave background (CMB) lensing potential derived from the \emph{Planck} nominal mission data and $z\gsim 1.5$ galaxies detected by the \emph{Herschel}-ATLAS  (H-ATLAS) survey covering about $600\,\hbox{deg}^2$, i.e. about 1.4\% of the sky. We reject the hypothesis that
there is no correlation between CMB lensing and galaxy detection at a $20\sigma$ significance, checking the result by performing a number of null tests. The significance of the detection of the theoretically expected cross-correlation is found to be $10\,\sigma$. The galaxy bias parameter, $b$, derived from a joint analysis of the cross-power spectrum and of the auto-power spectrum of the galaxy density contrast is found to be $b=2.80^{+0.12}_{-0.11}$, consistent with earlier estimates for H-ATLAS galaxies at similar redshifts.On the other hand, the amplitude of the cross-correlation is found to be a factor $A=1.62 \pm 0.16$ higher than expected from the standard model and also found by cross-correlation analyses with other tracers of the large-scale structure. The enhancement due to lensing magnification can account for only a fraction of the excess cross-correlation signal. We suggest that part of it may be due to an incomplete removal of the contamination of the cosmic infrared background, which includes the H-ATLAS sources we are cross-correlating with. In any case, the highly significant detection reported here using a catalog covering only 1.4\% of the sky demonstrates the potential of CMB lensing correlations with submillimeter surveys.
\end{abstract}
\keywords{galaxies: high-redshift, cosmic background radiation, gravitational lensing: weak, methods: data analysis, cosmology: observations}


\section{Introduction}

Cosmological observations carried out in the last two decades have enabled the establishment of the standard cosmological model. In this picture, observed galaxies form in matter overdensities that are the result of the growth, driven by gravitational instabilities in an expanding Universe, of primordial inhomogeneities generated during an inflationary epoch. A picture of primordial inhomogeneities at an early stage of their evolution is provided by observations of the cosmic microwave background (CMB) anisotropy.

However, this picture is to some extent distorted by interactions of the CMB photons with matter inhomogeneities encountered during their travel from the last-scattering surface to the observer. On the other hand, these effects are a useful source of information on the large-scale structure of the Universe. One of these effects is gravitational lensing, causing small but coherent deflections of the observed CMB temperature and polarization anisotropies, with a typical amplitude of $2'$. Specific statistical signatures of lensing enable the reconstruction of the gravitational potential integrated along the line of sight from observed CMB maps  \citep{hu:2002,hirata:2003}.

In recent years, CMB lensing has been measured in a number of CMB experiments. The first detections were made via cross-correlations with large-scale structures probed by galaxy surveys \citep{smith:2007,hirata:2008,feng:2012,bleem:2012,sherwin:2012,geach:2013}. The higher sensitivity and resolution of recent CMB instruments, such as the Atacama Cosmology Telescope (ACT), the South Pole Telscope (SPT), and \emph{Planck}, have enabled an internal detection of lensing using CMB data alone \citep{das:2011,keisler:2011,das:2013,engelen:2012}; the measurement with the highest signal-to-noise ratio (S/N), around 25$\sigma$, was reported last year by the \emph{Planck} team \citep{planck_lens:2013}.

\begin{figure} 
\plotone{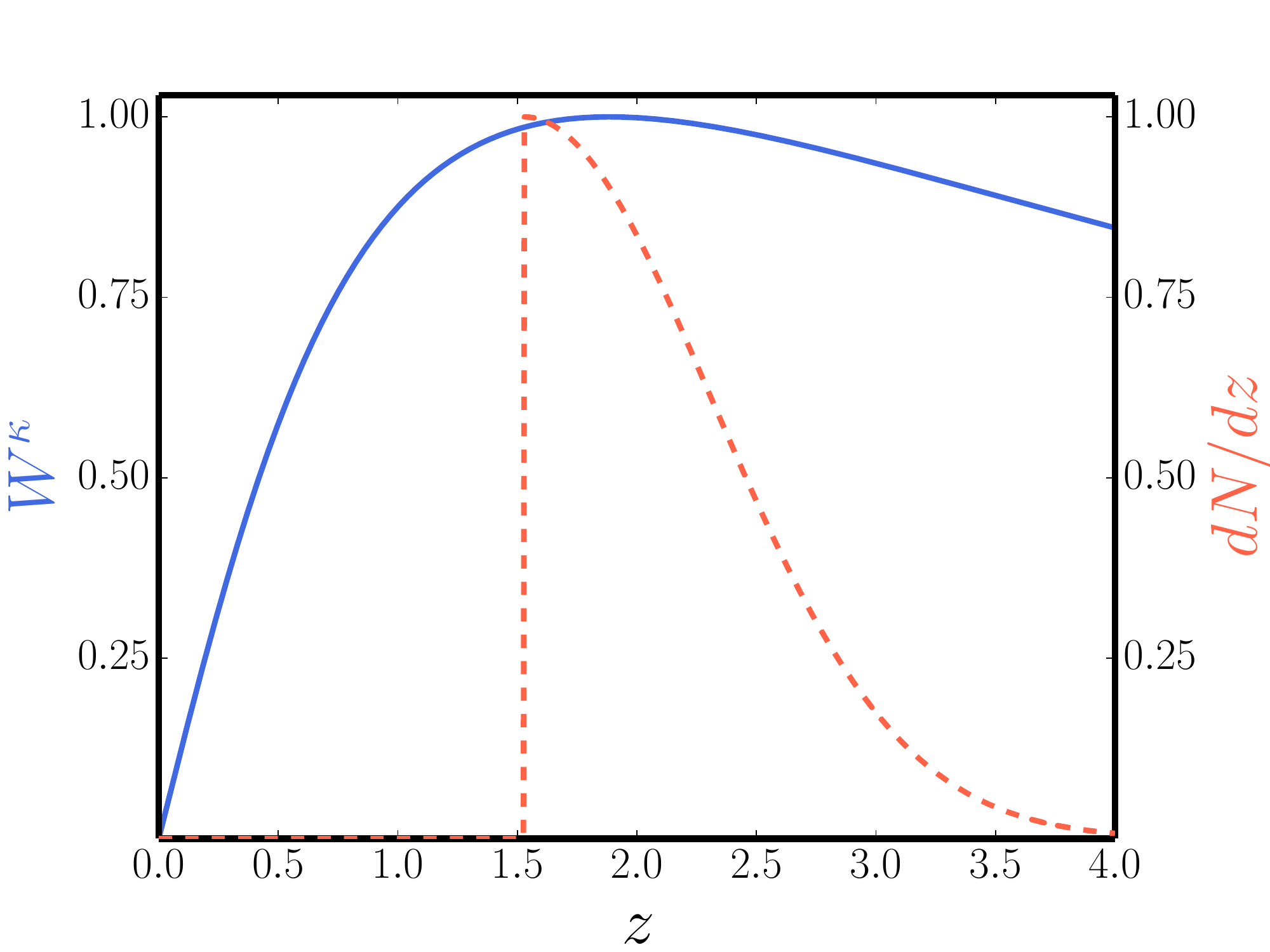}
\caption{Estimated redshift distribution of the full sample of H-ATLAS galaxies (dashed red line) compared with the CMB lensing kernel $W^{\kappa}$ (blue solid line). Both the kernels are normalized to a unit maximum. \label{fig:kernels_norm}}
\end{figure}

As already mentioned, the CMB lensing potential is an \textit{integrated} measure of the matter distribution in the Universe, up to the last-scattering surface. As illustrated by Figure~\ref{fig:kernels_norm}, it has a broad kernel, peaking at $z\simeq 2$ but slowly varying from $z\simeq 1$ to $z\gsim 4$. The study of cross-correlations with other tracers of large-scale structure covering narrow redshift ranges allows us to reconstruct the dynamics and spatial distribution of the 
cosmological gravitational potentials. This can tighten tests of the time evolution of dark matter density fluctuations and through that give constraints on the dynamics of the dark energy at the onset of cosmic acceleration. Because the cross-correlations measure the average lensing signal from the dark matter halos that host the galaxies, we can also derive from them the cosmic bias and hence the effective halo masses associated with the tracer populations. Although the bias factors can also be well determined from the autopower spectra, we must always beware of unaccounted systematic effects. The cross-correlation measurements are not prone to systematics that are not correlated between the two data sets. Thus a comparison of the bias estimates from auto- and cross-correlations can uncover unforeseen systematics on either side.

Several catalogs, such as those from the NRAO VLA Sky Survey (NVSS),  the Sloan Digital Sky Survey (SDSS), the Wide Field Survey Infrared Explorer (WISE) have already been cross-correlated with the CMB lensing potential. These surveys cover large areas of the sky but detected sources are mostly at $z\lsim 1$. The \textit{Herschel} Terahertz Large Area survey \citep[H-ATLAS;][]{eales10} allows us to extend the cross-correlation analysis up to substantially higher redshifts \citep{lapi11,gnuevo12}.

In this paper we present the first investigation of the cross-correlation between the CMB lensing potential measured by \textit{Planck} and \textit{Herschel}-selected galaxies with estimated redshifts $z\gsim 1.5$, i.e. at redshifts higher and closer to the peak of the lensing potential kernel than those of  the source samples considered so far. Our choice of restricting the analysis to $z\gsim 1.5$ has a twofold motivation. First, because we aim to reconstruct the evolution of the lensing potential at higher redshifts than done with other galaxy samples, it is expedient to remove the dilution of the signal by low-$z$ sources. Second, as shown by \citet{lapi11} and \citet{gnuevo12}, the adopted approach for estimating photometric redshifts becomes unreliable at $z \lsim 1$.

Highly statistically significant correlations between the CMB lensing and the cosmic infrared background (CIB) have been recently reported \citep{holder:2013,hanson:2013,planck_cib:2013,polarberar_herschel:2014}. There are obvious connections between these studies and the present one. However, the CIB is an integrated quantity and the interpretation of the measured cross-correlations depend on the adopted redshift distribution of sources, derived from a model. Our study of the cross-correlation with individually detected sources has the double advantage that redshifts are estimated directly from the data and are distributed over a quite narrow range.

The outline of this paper is as follows. In Section \ref{sec:theory} we describe the theoretical background while the data are introduced in Section \ref{sec:data}. The estimator of the cross-correlation power spectrum and the simulations used for validation of the algorithm and the error estimation are presented in Section \ref{sec:estimator}. The measured auto- and cross-power spectra, as well as the null tests, are reported in Section \ref{sec:power_spectra}. In Section \ref{sec:constraints}  we analyze the constraints on the galaxy bias and in Section \ref{sec:discussion} we discuss the potential systematic effects that affect the cross-correlation. Finally in Section \ref{sec:conclusions} we summarize our results.

Throughout this paper we adopt the fiducial flat $\Lambda$CDM cosmology with best-fit \emph{Planck} + WP + highL + lensing cosmological parameters as provided by the \emph{Planck} team in \cite{planck_parameters:2013}. Here, WP refers to WMAP polarization data at low multipoles, highL refers to the inclusion of high-resolution CMB data of the Atacama Cosmology Telescope (ACT) and South Pole Telescope (SPT) experiments, and lensing refers to the inclusion of \emph{Planck} CMB lensing data in the parameter likelihood.

\section{Theoretical Background}
\label{sec:theory}
The effect of gravitational lensing on CMB photons can be described as a remapping of the unlensed temperature anisotropies $\Theta(\nver)$ by a two-dimensional vector field in the sky, namely the deflection field $\mathbf{d}(\nver)$ \citep{lewis06}:
\begin{equation}
\begin{split}
\tilde{\Theta}(\hat{\mathbf{n}}) &= \Theta(\nver + \mathbf{d}(\nver)) \\
&= \Theta(\nver + \nabla\phi(\nver)) \\
&= \Theta(\nver) + \nabla^i \phi(\nver) \nabla_i \Theta(\nver) + \mathcal{O}(\phi^2),
\end{split}
\end{equation}
where $\tilde{\Theta}(\hat{\mathbf{n}})$ are the lensed temperature anisotropies and $\phi(\nver)$ is the CMB lensing potential:
\begin{equation}
\phi(\nver) = -2 \int_0^{z_*} \frac{c\,dz}{H(z)}\frac{\chi_* - \chi(z)}{\chi_*\chi(z)}\Psi(\chi(z)\nver,z).
\end{equation}
In this equation, $\chi(z)$ is the comoving distance to redshift $z$, $\chi_*$ is the comoving distance to the last-scattering surface at $z_*\simeq 1090$, $H(z)$ is the Hubble factor at redshift $z$, $c$ is the speed of light, and $\Psi(\chi(z)\nver,z)$ is the three-dimensional gravitational potential at a point on the photon path given by $\chi(z)\nver$. Note that the deflection angle is given by $\mathbf{d}(\nver) = \nabla\phi(\nver)$, where $\nabla$ is the the two-dimensional gradient on the sphere. Because the lensing potential is an integrated measure of the projected gravitational potential, taking the two-dimensional Laplacian of the lensing potential we can define the lensing convergence $\kappa(\nver) = -\case{1}{2}\nabla^2\phi(\nver)$, which depends on the projected matter overdensity $\delta$ \citep{bart01}:
\begin{equation}
\kappa(\nver) = \int_0^{z_*} dz\, W^{\kappa}(z)\delta(\chi(z)\nver,z).
\label{eqn:wkappa}
\end{equation}
The lensing kernel $W^{\kappa}$ is
\begin{equation}
W^{\kappa}(z) = \frac{3\Omega_{\rm m}}{2c}\frac{H_0^2}{H(z)}(1+z)\chi(z)\frac{\chi_*-\chi(z)}{\chi_*},
\end{equation}
where $\Omega_{\rm m}$ and $H_0$ are the present-day values of the Hubble and matter density parameters, respectively.

The galaxy overdensity $g(\nver)$ in a given direction on the sky is also expressed as a LOS integral of the matter overdensity:
\begin{equation}
g(\nver) = \int_0^{z_*} dz\, W^{g}(z)\delta(\chi(z)\nver,z),
\end{equation}
where the kernel is
\begin{equation}
\label{eqn:wg}
\begin{split}
W^{g}(z) &= \frac{b(z)\frac{dN}{dz}}{\Bigl(\int dz'\,\frac{dN}{dz'}\Bigr)} + \frac{3\Omega_{\rm m}}{2c}\frac{H_0^2}{H(z)}(1+z)\chi(z) \\
&\times \int_z^{z_*}dz'\,\Bigl(1-\frac{\chi(z)}{\chi(z')}\Bigr)(\alpha(z')-1)\frac{dN}{dz'}.
\end{split}
\end{equation}
The galaxy overdensity kernel is the sum of two terms: the first one is given by the product of the linear bias $b(z)$ and the redshift distribution $dN/dz$; and the second one takes into account the effect of gravitational magnification on the observed density of foreground sources \citep[magnification bias;][]{ho08,xia09}. This effect depends on the slope, $\alpha(z)$, of their integral counts ($N(>S) \propto S^{-\alpha}$) below the adopted flux density limit. Given the sharply peaked redshift distribution of our sources (see Figure~\ref{fig:kernels_norm}) we can safely assume a redshift- and scale-independent linear bias ($b(z)=\hbox{constant}$). Previous analyses of the clustering properties of submillimeter galaxies \citep{Xia2012,Cai2013} indicate $b\simeq 3$ at the redshifts of interest here, and we adopt this as our reference value.

Recent work by \cite{gonzalez-nuevo:2014} has shown that the magnification bias by weak lensing is substantial for high-$z$ H-ATLAS sources selected with the same criteria as the present sample (see the Section \ref{subsec:herschel}). This is because the source counts are steep, although their slope below the adopted flux density limit ($S_{250\mu\rm m}=35\,$mJy) is uncertain. The data \citep{Bethermin2012} indicate, at this limit, $\alpha \simeq 2$ while for the high-$z$ galaxy subsample considered in this work we find $\alpha \simeq 3$. In the following we adopt the latter as our fiducial value. The effect of different choices for this parameter value is examined in Section~\ref{sec:discussion}.

Because the relevant angular scales are much smaller than 1 radian (multipoles $\ell \ga 100$), the theoretical angular cross-correlation can be computed using the Limber approximation \citep{limber} as
\begin{equation}
C_{\ell}^{\kappa g} = \int_0^{z_*} \frac{dz}{c}\frac{H(z)}{\chi^2(z)}W^{\kappa}(z)W^{g}(z)P\Bigl(k=\frac{\ell}{\chi(z)},z\Bigr),
\label{eqn:kg}
\end{equation}
where $P(k,z)$ is the matter power spectrum, which we computed using the \textsc{CAMB}\footnote{available at \url{http://camb.info}} code \citep{camb}. The nonlinear evolution of the matter power spectrum was taken into account using the \textsc{HALOFIT} prescription \citep{halofit}. A more extended discussion on the effect of the nonlinear evolution in CMB lensing maps based on N-body simulations is carried out by \cite{antolini14}. The CMB convergence, $W^{\kappa}(z)$, and the galaxy redshift distribution $dN/dz$ of the sample analyzed in this work (see Section~\ref{subsec:herschel}) are shown in Figure \ref{fig:kernels_norm}.

\begin{figure} 
\plotone{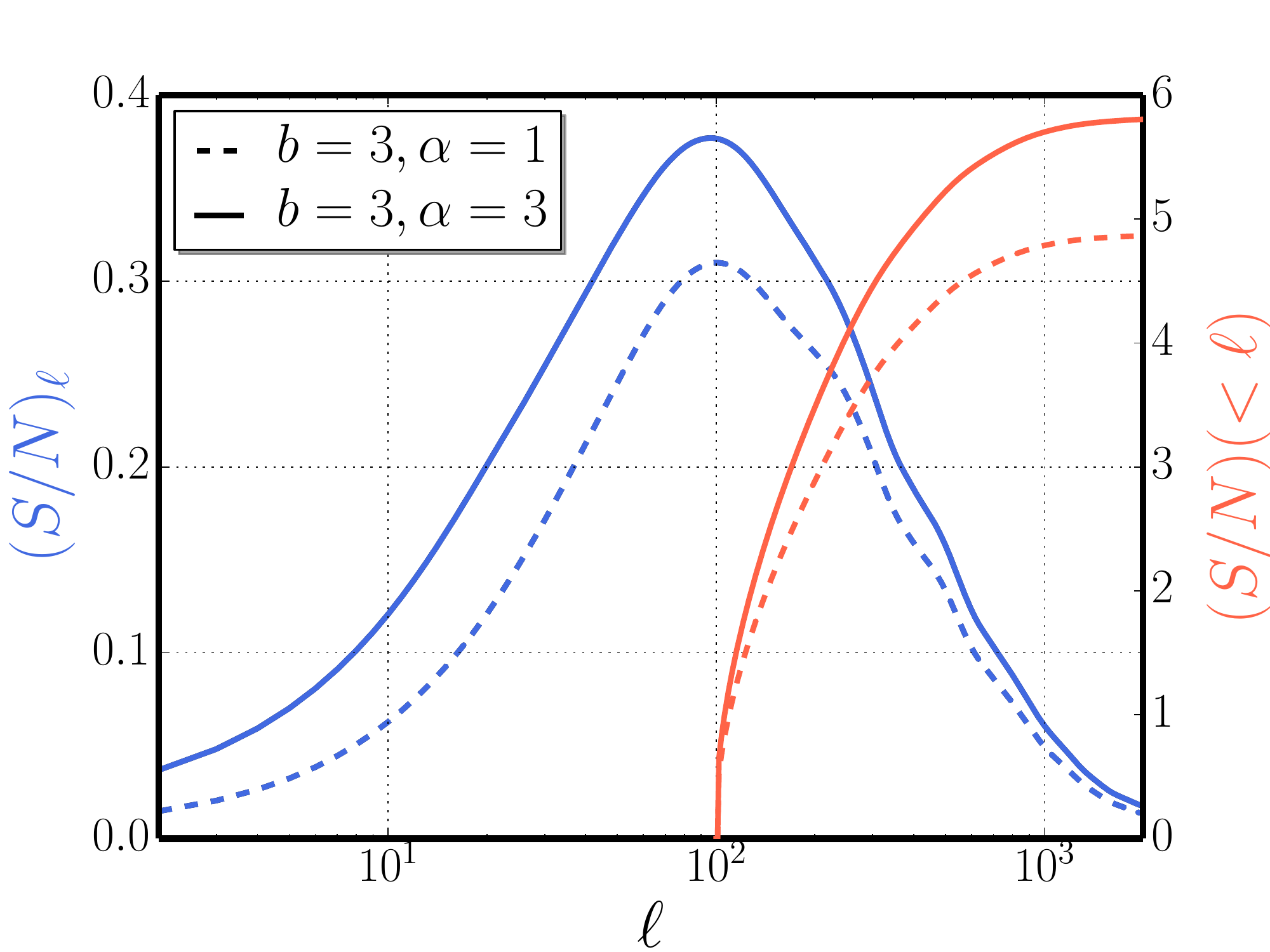}
\caption{S/N per multipole (blue lines; left axis) and cumulative S/N (red lines; right axis) evaluated from $\ell_{\rm min}=100$ for fiducial models with $b=3$ and $\alpha=1$ (no magnification, dashed lines) and $\alpha=3$ (solid lines). \label{fig:s_to_n_hatlas}}
\end{figure}

Again under the Limber approximation, the CMB convergence, $C_{\ell}^{\kappa\kappa}$, and the galaxy, $C_{\ell}^{gg}$, autospectra can be evaluated as
\begin{equation}
\begin{split}
C_{\ell}^{\kappa\kappa} &= \int_0^{z_*} \frac{dz}{c}\frac{H(z)}{\chi^2(z)}\Bigl[W^{\kappa}(z)\Bigr]^2P\Bigl(k=\frac{\ell}{\chi(z)},z\Bigr); \\
C_{\ell}^{gg} &= \int_0^{z_*} \frac{dz}{c}\frac{H(z)}{\chi^2(z)}\Bigl[W^{g}(z)\Bigr]^2P\Bigl(k=\frac{\ell}{\chi(z)},z\Bigr).
\end{split}
\label{eqn:kkgg}
\end{equation}
The mean redshift probed by the cross-correlation between CMB lensing and our sample is
\begin{equation}\label{eqn:meanz}
\langle z \rangle = \frac{\int_0^{z_*} \frac{dz}{c} z\frac{H(z)}{\chi^2(z)}W^{\kappa}(z)W^{g}(z)P\Bigl(k=\frac{\ell}{\chi(z)},z\Bigr)}{\int_0^{z_*} \frac{dz}{c}\frac{H(z)}{\chi^2(z)}W^{\kappa}(z)W^{g}(z)P\Bigl(k=\frac{\ell}{\chi(z)},z\Bigr)} \simeq 2.
\end{equation}
We can predict the S/N of the convergence-density correlation assuming that both the galaxy {overdensity} and the lensing fields behave as Gaussian random fields, so that the variance of $C_{\ell}^{\kappa g}$ is
\begin{equation}
\label{eqn:delta_kg}
\bigl(\Delta C_{\ell}^{\kappa g}\bigr)^2 = \frac{1}{(2\ell+1)f_{\rm sky}} \bigl[(C_{\ell}^{\kappa g})^2 + (C_{\ell}^{\kappa\kappa}+N_{\ell}^{\kappa\kappa})(C_{\ell}^{gg}+N_{\ell}^{gg})\bigr],
\end{equation}
where $f_{\rm sky}$ is the sky fraction covered by both the galaxy and the lensing surveys, $N_{\ell}^{\kappa\kappa}$ is the noise of the lensing field, and $N_{\ell}^{gg}=1/\bar{n}$ is the shot noise associated with the galaxy field. Because our calculations are done in terms of the density contrast, the shot noise is inversely proportional to the mean number of sources per steradian, $\bar{n}$. The signal to noise ratio at multipole $\ell$ is then
\begin{equation}
\label{eqn:s_to_n_mult}
\Bigl( \frac{S}{N} \Bigr)^2_{\ell} = \frac{\bigl(C_{\ell}^{\kappa g}\bigr)^2}{\bigl(\Delta C_{\ell}^{\kappa g}\bigr)^2} = \frac{(2\ell+1)f_{\rm sky}\bigl(C_{\ell}^{\kappa g}\bigr)^2 }{\bigl[(C_{\ell}^{\kappa g})^2 + (C_{\ell}^{\kappa\kappa}+N_{\ell}^{\kappa\kappa})(C_{\ell}^{gg}+N_{\ell}^{gg})\bigr]},
\end{equation}
and the cumulative S/N for multipoles up to $\ell_{\rm max}$ is
\begin{equation}
\label{eqn:s_to_n_cum}
\Bigl( \frac{S}{N} \Bigr)(<\ell_{\rm max}) = \sqrt{\sum_{\ell'=\ell_{\rm min}}^{\ell_{\rm max}} \Bigl( \frac{S}{N} \Bigr)^2_{\ell'}}.
\end{equation}
In Figure \ref{fig:s_to_n_hatlas} we show both the S/N per multipole and the cumulative one computed using the specifications for the \textit{Planck} lensing noise (see Section ~\ref{subsec:planck}) and the mean surface density of our source sample. It must be noted that, because of the limited area covered by the H-ATLAS survey (and split into 5 fields), the cross-correlation is only meaningful on scales below a few degrees. We have therefore limited our analysis to $\ell \ge \ell_{\rm min} = 100$. This restriction prevents us from exploiting the peak at $\ell \sim 100$ of the signal to noise per multipole. The cumulative S/N saturates at $\ell \sim 1000$. If $b=3$ and $\alpha=3$ we expect $S/N \simeq 6$.

\section{Data}
\label{sec:data}
\subsection{\emph{Planck} data}
\label{subsec:planck}
We used the publicly released \emph{Planck} CMB lensing potential map derived from the first 15.5 months of observations \citep{planck_lens:2013}.  The \emph{Planck} satellite observed the sky with high angular resolution in nine frequency bands, from 30 to 857 GHz \citep{planck_general:2013}. The angular resolution ($10'$, $7'$, and $5'$) and the noise level (105, 45 and 60 $\mu$K\,arcmin) of the 100, 143 and 217 GHz frequency channels, respectively, make them the most suitable for estimation of the gravitational lensing potential. Nevertheless, the released map is based on a minimum variance combination of the 143 and 217 GHz temperature anisotropy maps only, because adding the 100 GHz map yields a negligible improvement \citep{planck_lens:2013}. The maps are in the HEALPix\footnote{\url{http://healpix.jpl.nasa.gov}} \citep{healpix} format with a resolution parameter of $N_{\rm side} = 2048$, corresponding to 50, 331, and 648 pixels over the sky, with a pixel size of $\sim 1.7'$.

\begin{figure} 
\plotone{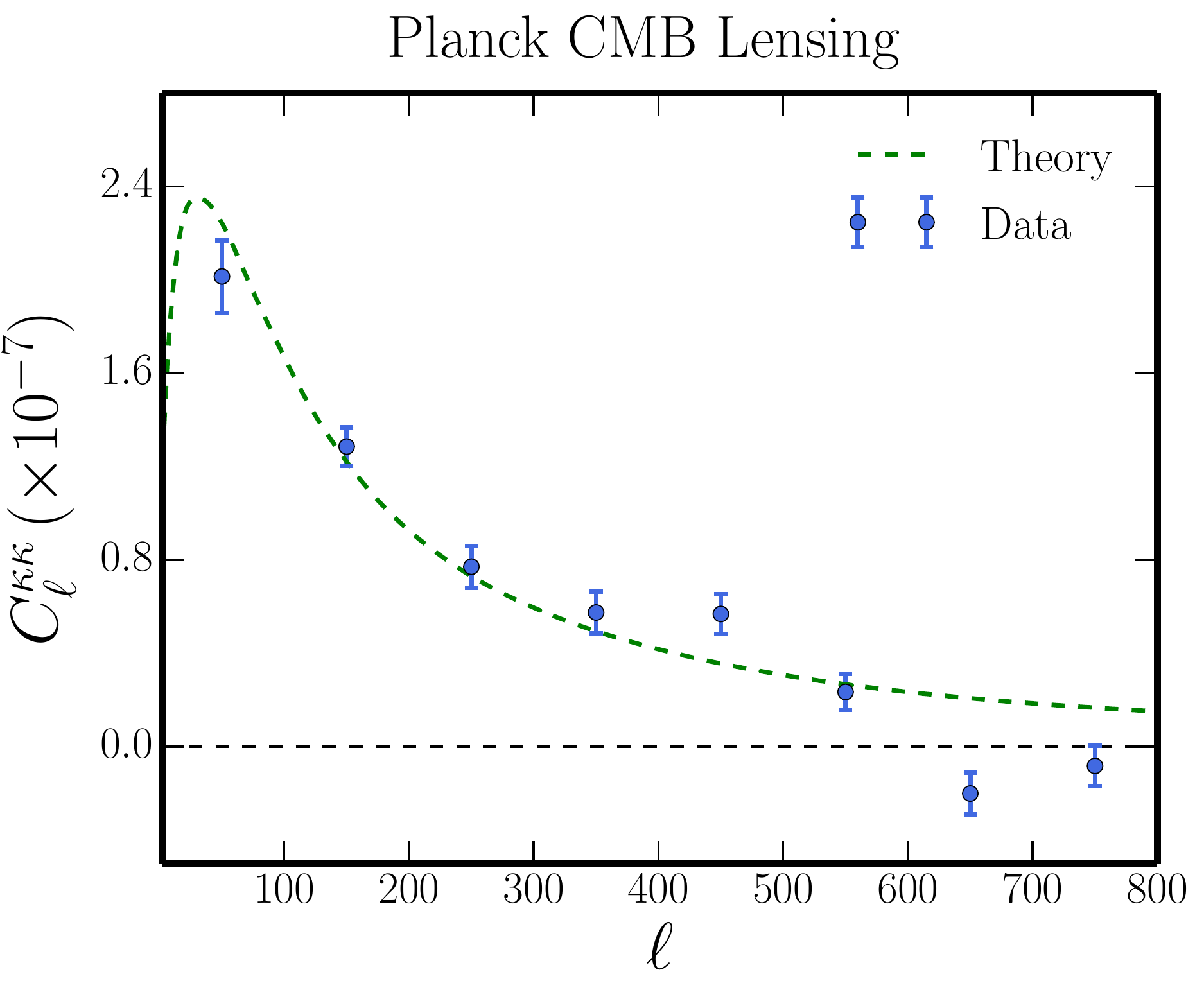}
\caption{CMB convergence autopower spectrum as reconstructed from \emph{Planck} data (blue points) on a portion of the sky with $f_{\rm sky} \simeq 0.6$ compared with the theoretical prediction for our background cosmology (dashed green line).
 \label{fig:kk_data_planck}}
\end{figure}

The power spectrum of the lensing potential is very red, and this may introduce a bias when we estimate it within multipole bins. To avoid this problem, we decided to convert the lensing potential map, $\phi$, into the convergence map, $\kappa$, which has a much less red power spectrum.
This was done using the relation between the spherical harmonic coefficients of these quantities estimated on the full sky \citep{hu00}
\begin{equation}
\kappa_{\ell m} = -\frac{\ell(\ell+1)}{2}\phi_{\ell m} \ .
\end{equation}
The convergence spherical harmonic coefficients were transformed to a map with resolution parameter $N_{\rm side}=512$ corresponding to a pixel size of $\sim 7'$. This resolution is sufficient for our analysis because the data noise level enables us to detect cross-correlations between the convergence and the galaxy density field only for angular scales larger than $\sim 20'$ ($\ell \lesssim 540$).

The convergence autopower spectrum recovered on approximately $60\%$ of the sky using a modified version of the mask provided by the \textit{Planck} collaboration is shown in Figure~\ref{fig:kk_data_planck}. The auto-power spectrum has been corrected for the lensing reconstruction noise power spectrum  $N_{\ell}^{\kappa\kappa}$ which was estimated from the set of 100 simulated lensing maps\footnote{\url{http://irsa.ipac.caltech.edu/data/Planck/release_1/ancillary-data/HFI_Products.html}} recently released by the \emph{Planck} team that account for the inhomogeneous noise level. The noise power spectrum was computed by averaging the spectra of the difference maps between the reconstructed and the input lensing map over 100 realizations. The errors on band powers were calculated as the diagonal part of the covariance matrix built from the simulation, as described in Section.~\ref{sec:estimator}. The raw auto-power spectrum is not corrected for the bias induced by non-Gaussianity of unresolved point sources and for pseudo-$C_{\ell}$ leakage effects from masking (we just correct for N0 and N1 bias term adopting the formalism of \cite{planck_lens:2013}). These terms may cause some discrepancy of the power spectrum at high multipoles. Nevertheless, in the range of multipoles relevant for our analysis the power spectrum agrees pretty well with the theoretical one, and proper estimation of the convergence power spectrum is outside the scope of this paper.

\subsection{\textit{Herschel} fields}
\label{subsec:herschel}

We exploited the data collected by the \textit{Herschel} Space Observatory \citep{pilbratt10} in the context of the \textit{Herschel} Astrophysical Terahertz Large Area Survey \citep[H-ATLAS;][]{eales10}, an open-time key program that has surveyed about $550$ deg$^2$ with the Photodetector Array Camera and Spectrometer \citep[PACS;][]{poglitsch10} and the Spectral and Photometric Imaging Receiver \citep[SPIRE;][]{griffin10} in five bands, from $100$ to $500\,\mu$m. The H-ATLAS mapmaking is described by \citet{Pascale2011} for SPIRE and by \citet{Ibar2010} for PACS. The procedures for source extraction and catalog generation can be found in \citet{Rigby2011}, Maddox \textit{et al.} (2015, in preparation) and Valiante \textit{et al.} (2015, in preparation).

The survey area is divided into five fields: three equatorial fields centered on 9hr, 12hr, and 14.5hr (GAMA fields, G09, G12, and G15) covering, altogether, $161\,\hbox{deg}^2$; the north galactic pole (NGP) block, a rectangular block of $15^\circ\,\cos(\delta)$ by $10^\circ$ centered on right ascension $\alpha=199.5^\circ$, declination $\delta=29^\circ$ and rotated by approximately $8^\circ$ clockwise; and the south galactic pole (SGP) block consisting of two concatenated rectangular regions, one of $31.5^\circ\cos(\delta)$ by $6^\circ$ centered on $\alpha=351.3^\circ$, $\delta=-32.8^\circ$, the other of $20^\circ\cos(\delta)$ by $6^\circ$ centered on $\alpha=18.1^\circ$, $\delta=-30.7^\circ$.

The $z\lsim 1$ galaxies detected by the H-ATLAS survey are mostly late-type and starburst galaxies with moderate star-formation rates and relatively weak clustering  \citep{dunne11,guo11}. High-$z$ galaxies are forming stars at high rates ($\ge \hbox{few hundred}\,\hbox{M}_\odot\,\hbox{yr}^{-1}$) and are much more strongly clustered \citep{Maddox2010,Xia2012}, implying that they are tracers of large-scale overdensities. Their properties are consistent with them being the progenitors of local massive elliptical galaxies \citep{lapi11}. We aim to correlate high-$z$ H-ATLAS galaxies with the \textit{Planck} CMB lensing map.

To select the high-$z$ population, we adopted the criteria developed by \cite{gnuevo12}: (i) $S_{250\,\mu\rm m}>35$ mJy; (ii) $S_{350\,\mu\rm m}/S_{250\,\mu\rm m}>0.6$ and $S_{500\,\mu\rm m}/S_{350\,\mu\rm m}>0.4$ ; (iii) $3\,\sigma$ detection at $350\,\mu$m; and (iv) photometric redshift $z_{\rm phot}>1.5$, estimated following \citet{lapi11} and \citet{gnuevo12}. 

Our final sample comprises a total of 99,823 sources, of which $9,099$ are in G09, $8,751$ in G12, $9,279$ in G15, $28,245$ in NGP and $44,449$ in SGP. The specifics of each patch are summarized in Table ~\ref{herschel_patches}. The redshift distribution of the population is needed in order to predict the amplitude of the cross-correlation. Estimating the uncertainties in the redshift distribution due to photometric redshift errors is not a trivial task. 

\begin{figure} 
\plotone{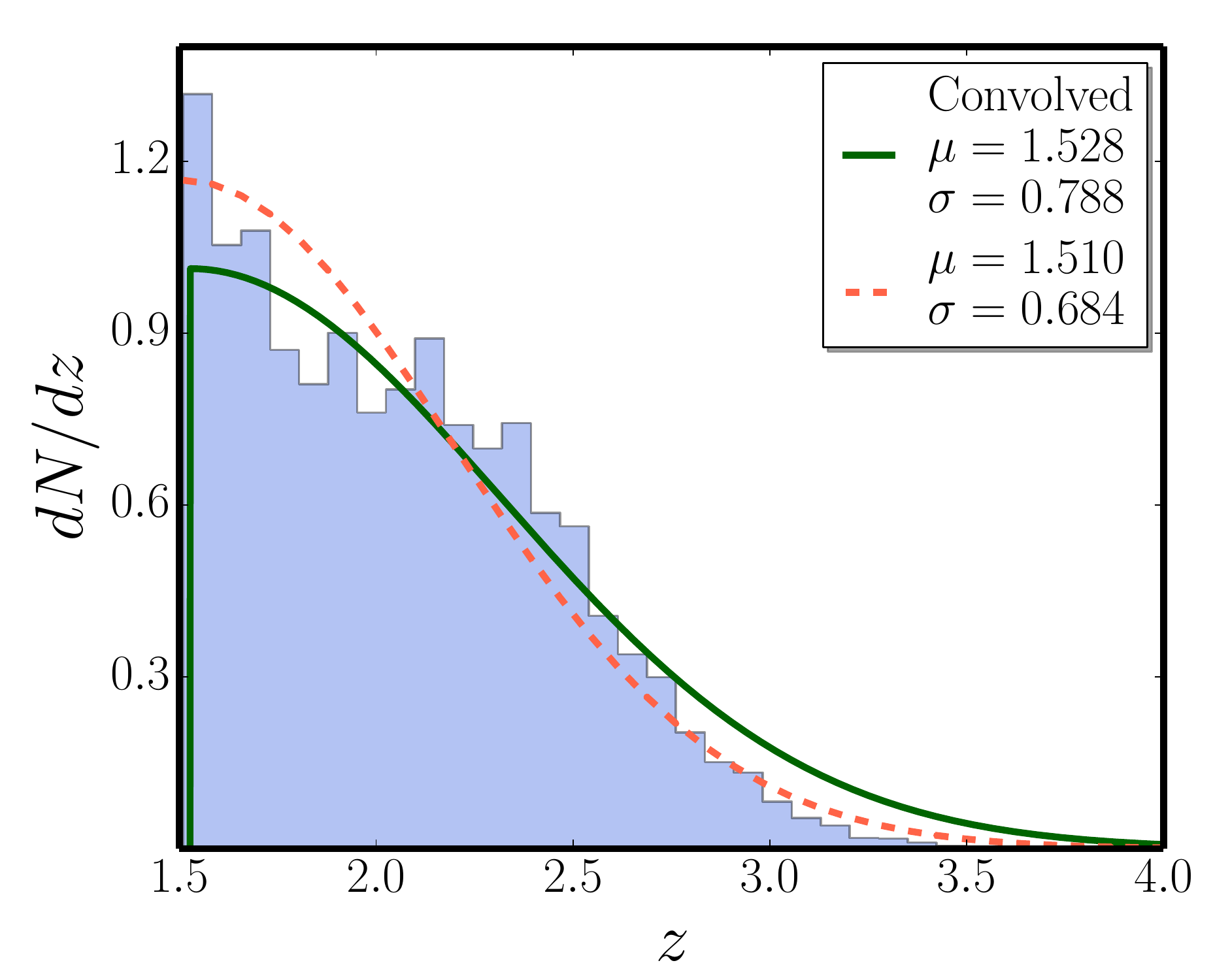}
\caption{Redshift distribution of H-ATLAS galaxies for the combined set of patches used in the analysis. The (blue) histogram  is the empirical redshift distributions, the dashed (orange) line is the half-normal fit to $dN/dz$ as described in text, while the solid (green) line represents the convolved $dN/dz$ that takes into account errors on photo-z estimation and is used as the fiducial distribution in our analysis. The values of the parameters $\mu$ and $\sigma$ given in the box are the best-fit values and are used in the analytic expression for $dN/dz$ adopted in calculations. \label{fig:dndz}}
\end{figure}

As stated in \citet{gnuevo12} there is no indication that photometric redshifts are systematically under- or overestimated when the spectral energy distribution of SMM J2135-0102 is used as a template. The median value of $ \Delta z/(1 + z) \equiv (z_{\rm phot} -  z_{\rm spec})/(1 + z_{\rm spec})$ is -0.002 with a dispersion of 0.115. This dispersion corresponds to an rms error on $z$ of $\sigma_{\langle z \rangle}=  0.345$ at the mean redshift $\langle z \rangle\simeq 2$, given by Equation~(\ref{eqn:meanz}). To get a rough indication of how many sources were scattered above and below the redshift threshold ($z=1.5)$ by measurement errors we have convolved a gaussian fit to the redshift distribution of sources selected with the first 3 criteria [(1) to (3)] with a gaussian error distribution having zero mean and dispersion $\sigma_{\langle z \rangle}$. The convolved redshift distribution was cut at $z = 1.5$, and the portion at higher $z$ was fitted with a half-normal distribution normalized to unity:
\begin{equation}
\frac{dN}{dz} = \frac{\sqrt{2}}{\sigma\sqrt{\pi}}\exp{\Bigl( -\frac{(z-\mu)^2}{2\sigma^2}\Bigr)}.
\end{equation}
The redshift distributions of the galaxies before and after the convolution are shown in Figure~\ref{fig:dndz}.

\begin{figure*} 
\plotone{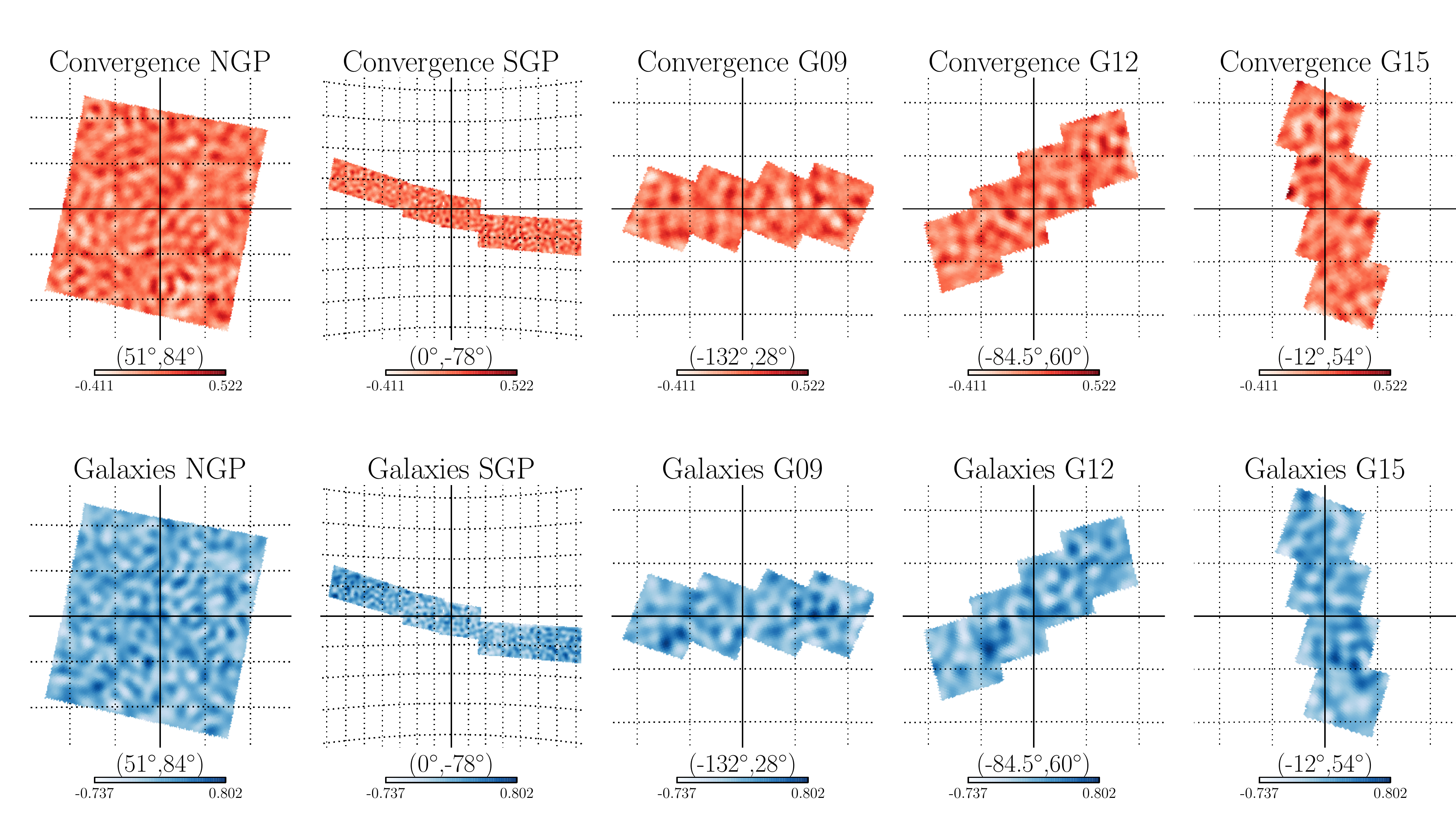}
\caption{Convergence maps (upper row) and galaxy overdensity maps (lower row) in the H-ATLAS fields: multipoles $\ell > 400$ for which $(S/N)_{\ell} \lesssim 0.3$ have been filtered out. Galactic longitude and latitude $(l,b)$ of patch centers are provided in brackets. The grid overlay has spacing of $3^\circ$ in each box. \label{fig:patches}}
\end{figure*}

We built an overdensity map at a resolution $N_{\rm side}=512$ defined by
\begin{equation}
\label{eqn:counttodensity}
g(\nver) = \frac{n(\nver)-\bar{n}}{\bar{n}},
\end{equation}
where $n(\nver)$ is the number of objects in a given pixel, and $\bar{n}$ is the mean number of objects per pixel. The CMB convergence and galaxy overdensity maps in the different patches are shown in Figure~\ref{fig:patches}. We filtered out from these fields multipoles $\ell \gtrsim 400$ where $(S/N)_{\ell}\lesssim 0.3$.

\begin{deluxetable}{ccccc}
\tabletypesize{}
\tablecaption{H-ATLAS Patches Data\label{herschel_patches}}
\tablewidth{0pt}
\tablehead{
\colhead{Patch} & \colhead{$N_{\rm obj}$} & \colhead{$f_{\rm sky}$} & \colhead{$\bar{n}\,[\hbox{gal}\,\hbox{pix}^{-1}]$} & \colhead{$\bar{n}\,[\hbox{gal}\,\hbox{sr}^{-1}]$}
}
\startdata
ALL   &   99823 &   0.014   &    2.30    &    $5.76\times 10^5$ \\
NGP  &   28245 &   0.004   &    2.25    &    $5.64\times 10^5$ \\
SGP  &   44449 &   0.006   &    2.38    &    $5.95\times 10^5$ \\
G09   &   9099   &   0.001   &    2.28    &    $5.71\times 10^5$ \\
G12   &   8751   &   0.001   &    2.13    &    $5.35\times 10^5$ \\
G15   &   9279   &   0.001   &    2.27    &    $5.68\times 10^5$ \\
\enddata
\tablenotetext{a}{ALL is  the combination of all the patches together.}
\end{deluxetable}

\section{The Cross-Correlation Algorithm}
\label{sec:estimator}

\subsection{Estimator}

We computed the angular power spectra within the regions covered by the H-ATLAS survey using a pseudo-$C_\ell$ estimator based on the MASTER algorithm \citep{master}. These regions are inside the area used in the estimation of the CMB lensing map. For a survey that covers only a fraction of the sky, different modes of the true cross-power spectrum $C^{\kappa g}_{\ell}$ are coupled \citep{peebles}. The coupling can be described by the mode-mode coupling matrix $M_{\ell\ell'}$ which relates the pseudo-cross-spectrum $\tilde{C}^{\kappa g}_{\ell}$ measured from the data
\begin{equation}
\tilde{C}^{\kappa g}_{\ell} = \frac{1}{2\ell+1}\sum_{m=-\ell}^{\ell} \tilde{\kappa}_{\ell m}\tilde{g}^*_{\ell m}.
\end{equation}
to the true spectrum
\begin{equation}
\label{eqn:coupling}
\tilde{C}_{\ell}^{\kappa g} = \sum_{\ell'} M_{\ell\ell'}C_{\ell'}^{\kappa g}.
\end{equation}
However, we cannot directly invert Equation~\eqref{eqn:coupling} to get the true power spectrum, because for surveys covering only a small fraction of the sky, the coupling matrix $M_{\ell\ell'}$ becomes singular.  To reduce the correlations of the $C_{\ell}$'s it is necessary to bin the power spectrum in $\ell$. We used eight linearly spaced bins of width $\Delta\ell = 100$ in the range $0 \le \ell \le 800$.

Then, the estimator of the true band powers $\hat{C}^{\kappa g}_{L}$ (hereafter $C^{\kappa g}_{L}$ denotes the binned power spectrum and  $L$ identifies the bin) is given by
\begin{equation}
\label{eqn:master_kg}
\hat{C}^{\kappa g}_{L} = \sum_{L' \ell}K^{-1}_{LL'}P_{L'\ell}\tilde{C}^{\kappa g}_{\ell},
\end{equation}
where
\begin{equation}
K_{LL'} = \sum_{\ell\ell'} P_{L\ell}M_{\ell\ell'}B^2_{\ell'}Q_{\ell' L'}.
\end{equation}
Here $P_{L\ell}$ is the binning operator; $Q_{\ell L}$ and $B^2_{\ell'}$ are, respectively, the reciprocal of the binning operator and  the pixel window function that takes into account the finite pixel size. Because of the small size of the sky area covered by the H-ATLAS survey, the power spectrum for $\ell < 100$ is very poorly estimated, and we did not use it in our analysis. However, to avoid the bias coming from the lowest-order multipoles, the first multipole bin is included in the computation of the power spectrum; that is, the inversion of the binned coupling matrix $K_{LL'}$ is performed including the first bin, and the pseudopower spectrum for the first bin is used in the product of Equation~\eqref{eqn:master_kg}.

The main assumption in cross-correlation studies is that the noise levels related to the observables being analyzed are uncorrelated, so that we do not need to debias the reconstructed cross-spectrum for any noise term. However, when dealing with autopower spectra, such as $C^{gg}_{\ell}$ and $C^{\kappa\kappa}_{\ell}$, we have to correct the estimator given by Equation~\eqref{eqn:master_kg} in order to account for the noise:
\begin{equation}
\begin{split}
\hat{C}^{gg}_{L} &= \sum_{L'\ell}K^{-1}_{LL'}P_{L'\ell}\Bigl( \tilde{C}^{gg}_{\ell} - \langle \tilde{N}^{gg}_{\ell}\rangle_{\rm MC} \Bigr), \\
\hat{C}^{\kappa\kappa}_{L} &= \sum_{L'\ell}K^{-1}_{LL'}P_{L'\ell}\Bigl( \tilde{C}^{\kappa\kappa}_{\ell} - \langle \tilde{N}^{\kappa\kappa}_{\ell}\rangle_{\rm MC} \Bigr),
\end{split}
\end{equation}
where $\langle \tilde{N}^{gg}_{\ell}\rangle_{\rm MC}$ and $\langle \tilde{N}^{\kappa\kappa}_{\ell}\rangle_{\rm MC}$ are the average noise pseudospectra estimated from the Monte Carlo (MC) simulations.

\begin{figure*}
\centering
\includegraphics[scale=0.28]{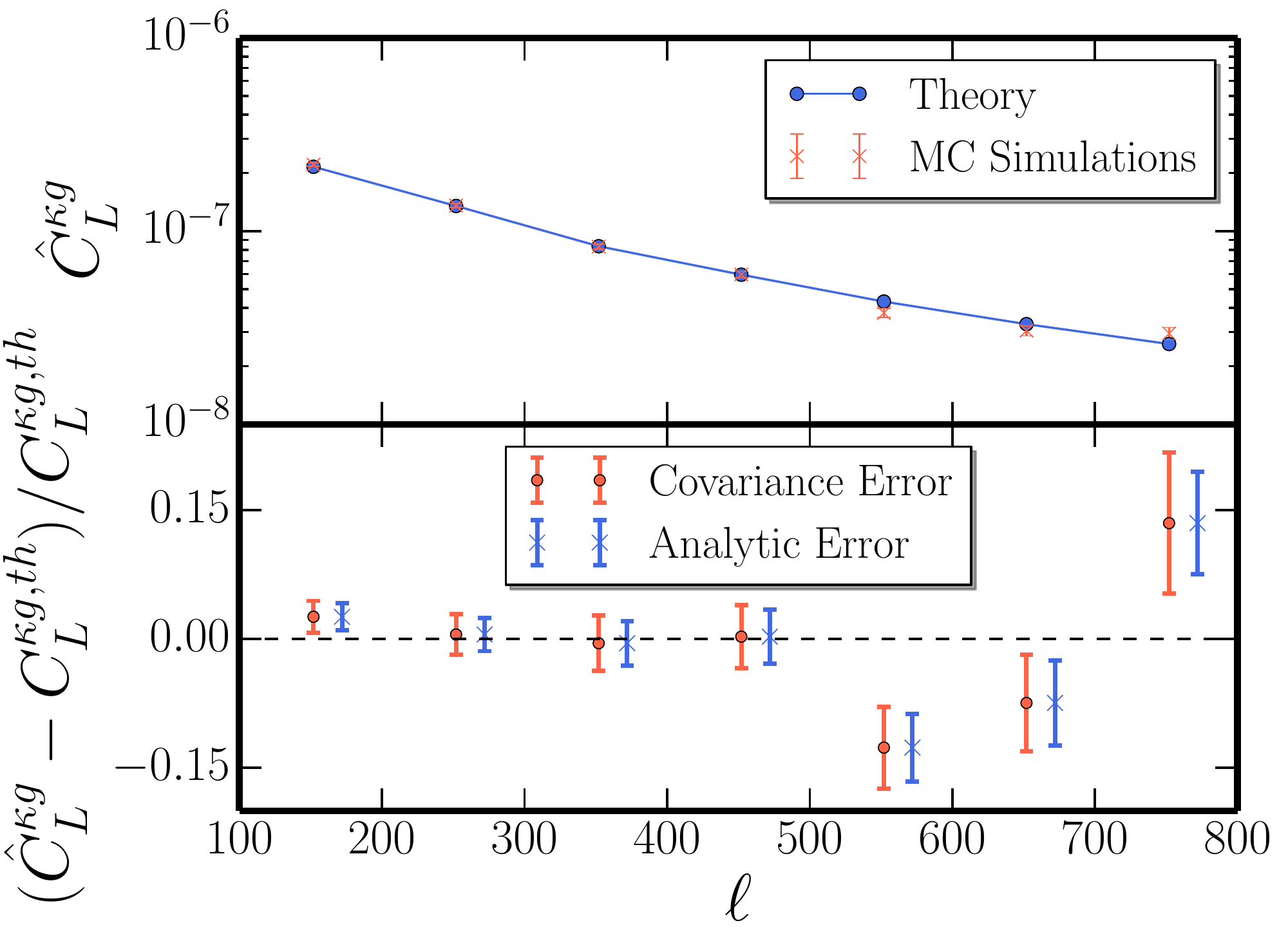}
\,
\includegraphics[scale=0.28]{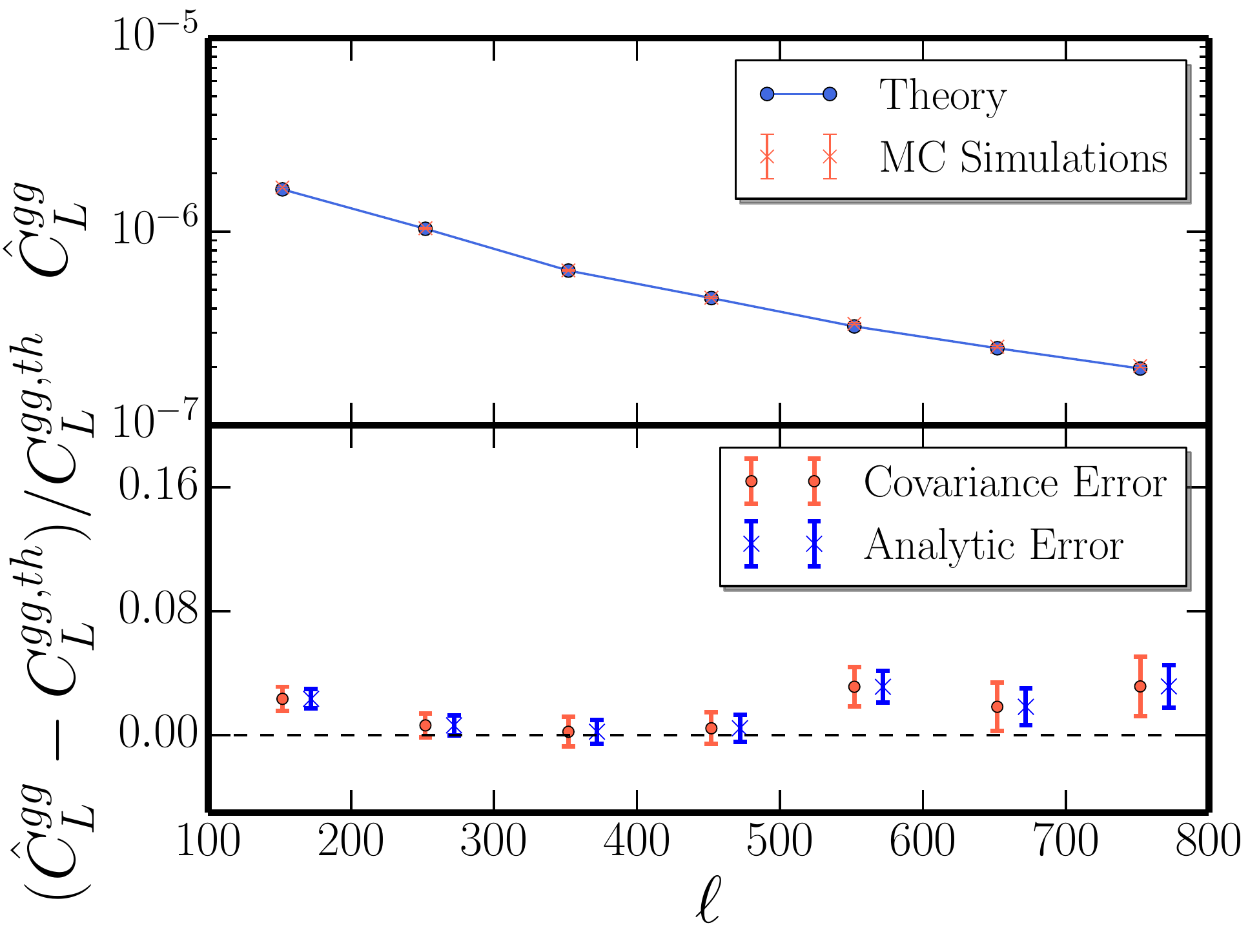}
\,
\includegraphics[scale=0.28]{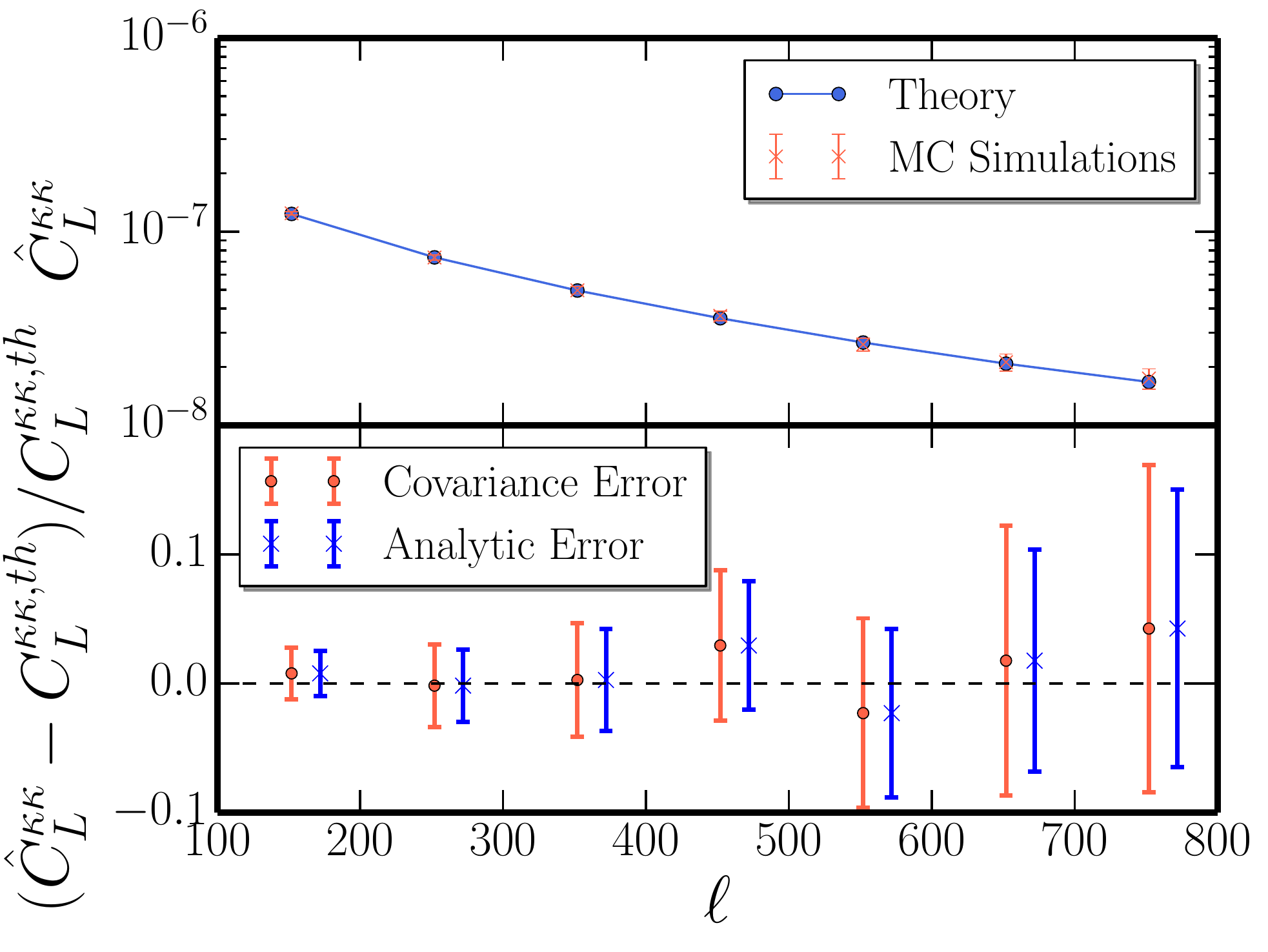}
\caption{\emph{Left}. \emph{Upper panel}: cross-power spectrum of simulated galaxy and lensing maps constructed with $b=3$. The points connected by the solid blue line represent the binned input cross-spectrum, and the average reconstructed spectrum from 500 simulations is shown by the orange points. \emph{Lower panel}:  fractional difference between the input and extracted cross-spectra. Error bars obtained with the simulation covariance matrix (orange points) and with the analytical approximation (blue points) are shown for comparison. \emph{Middle}. As in left plot, but for the galaxy auto-power spectrum. \emph{Right}. As in left plot, but for the CMB convergence autopower spectrum}
\label{fig:validation}
\end{figure*}

\subsection{Covariance matrix}\label{sec:cov}
The errors on the cross-power spectrum are described by the covariance matrix \citep{brown:2005}
\begin{equation}
\label{eqn:analytic_covariance}
 \text{Cov}^{\kappa g}_{LL'} = M_{LL_1}^{-1} P_{L_1\ell}\,\widetilde{\text{Cov}}^{\kappa g}_{\ell\ell'} \, Q_{\ell' L_2} (M_{L'L_2}^{-1})^T,
\end{equation}
where $\widetilde{\text{Cov}}^{\kappa g}_{\ell\ell'}$ is the pseudocovariance matrix given by
\begin{equation}
\begin{split}
\label{eqn:pcov_kg}
 &\widetilde{\text{Cov}}^{\kappa g}_{\ell\ell'}  =  \frac{1}{2\ell'+1}M_{\ell\ell'}\Bigl[ C_{\ell}^{\kappa g}(b)C_{\ell'}^{\kappa g}(b)+ \\
 & \sqrt{(C_{\ell}^{\kappa\kappa}+N_{\ell}^{\kappa\kappa})(C_{\ell}^{gg}(b) + N_{\ell'}^{gg})(C_{\ell'}^{\kappa\kappa}+N_{\ell'}^{\kappa\kappa})(C_{\ell'}^{gg}(b) + N_{\ell'}^{gg})}\Bigr].
\end{split}
\end{equation}
The corresponding covariance matrix of the galaxy autocorrelation is obtained by replacing in Equation~(\ref{eqn:analytic_covariance}) the pseudocovariance matrix $\widetilde{\text{Cov}}^{\kappa g}_{\ell\ell'}$ with  $\widetilde{\text{Cov}}^{gg}_{\ell\ell'}$ given by
\begin{equation}
\label{eqn:pcov_gg}
 \widetilde{\text{Cov}}^{gg}_{\ell\ell'} = \frac{2}{2\ell'+1}M_{\ell\ell'}\Bigl[(C_{\ell}^{gg}(b) + N_{\ell}^{gg})(C_{\ell'}^{gg}(b) + N_{\ell'}^{gg})\Bigr].
\end{equation}
The analytical expressions for the covariance matrices given above were used in the estimation of the galaxy bias and of the amplitude of the cross-correlation, presented in Section \ref{sec:constraints}.

\subsection{Validation}
In order to validate the algorithms used for the computation of the estimators outlined in the previous section and to check that the cross- and autopower spectra estimates are unbiased, we created 500 simulated maps of the CMB convergence field and of the galaxy overdensity field with statistical properties consistent with observations.

Using the theoretical spectra obtained with eqs. \eqref{eqn:kg} and \eqref{eqn:kkgg}, we generated full-sky signal maps, injecting a known degree of correlation, so that the simulated CMB convergence and galaxy harmonic modes satisfy both the auto- and the cross-correlations \citep{Kamionkowski1997}:
\begin{equation}
\begin{split}
\kappa_{\ell m} &= \zeta_1 \bigl(C_{\ell}^{\kappa\kappa} \bigr)^{1/2};\\
g_{\ell m} &= \zeta_1 \frac{C_{\ell}^{\kappa g}}{\bigl(C_{\ell}^{\kappa\kappa} \bigr)^{1/2}} + \zeta_2 \Biggl[ C_{\ell}^{gg} - \frac{\bigl(C_{\ell}^{\kappa g}\bigr)^2}{C_{\ell}^{\kappa\kappa}} \Biggr]^{1/2}.
\end{split}
\end{equation}
For each value of $\ell$ and $m>0$, $\zeta_1$ and $\zeta_2$ are two complex numbers drawn from a Gaussian distribution with unit variance, whereas for $m=0$ they are real and normally distributed.

We also generated 500 noise realizations for both fields. To simulate Gaussian convergence noise maps, we used the convergence noise power spectrum $N^{\kappa\kappa}_{\ell}$ provided by the \emph{Planck} team\footnote{\url{http://wiki.cosmos.esa.int/planckpla/index.php/Specially_processed_maps}}. Although this power spectrum is not sufficiently accurate to estimate the convergence power spectrum, as pointed out in the \emph{Planck} Collaboration Products Web site, it should be sufficiently good for the cross-correlation analysis, which is not biased by the noise term. For the same reason, it is not crucial for our analysis to use the 100 simulations of the estimated lensing maps provided recently by the \emph{Planck} team.

To take into account noise in the simulated galaxy maps, we proceeded in the following way. For each signal map containing the galaxy overdensity, we generated a set of  simulated galaxy number count maps, where the value in each pixel is drawn from a Poisson distribution with mean
\begin{equation}
\lambda(\nver) = \bar{n}(1+g(\nver)),
\end{equation}
where $\bar{n}$ is the mean number of sources per pixel in a given H-ATLAS patch and $g(\nver)$ is the corresponding simulated galaxy map containing only signal. The galaxy number counts map $\lambda(\nver)$ was then converted into a galaxy overdensity map using Equation~\eqref{eqn:counttodensity}, substituting the real number of objects in a given pixel $n(\nver)$ with the simulated one $\lambda(\nver)$. Note that maps obtained in this way already include Poisson noise with variance $N^{gg}_{\ell} = 1/\bar{n}$.

We applied the pipeline described above to our set of simulations in order to recover the input cross- and autopower spectra used to generate such simulations. The extracted $\hat{C}^{\kappa g}_{L}$, $\hat{C}^{gg}_{L}$, and $\hat{C}^{\kappa\kappa}_{L}$  spectra averaged over 500 simulations are reported in Figure~\ref{fig:validation}.  The mean band power was computed as
\begin{equation}
\langle \hat{C}^{XY}_{L} \rangle = \frac{1}{\Nsim}\sum_{i=1}^{\Nsim} \hat{C}^{XY,i}_{L},
\end{equation}
where $X,Y = \{\kappa,g\}$, $i$ refers to the $i$-th simulation, and $\Nsim=500$ is the number of simulations. The errors were computed from the covariance matrix as
\begin{equation}
\Delta \hat{C}^{XY}_{L} = \Bigl(\frac{\cov^{XY}_{LL}}{\Nsim} \Bigr)^{1/2},
\end{equation}
and the covariance matrix $\cov^{XY}_{LL'}$ was evaluated from the simulations as
\begin{equation}
\cov^{XY}_{LL'} = \frac{1}{\Nsim-1} \sum_{i=1}^{\Nsim} ( \hat{C}^{XY,i}_{L} - \langle \hat{C}^{XY}_{L}\rangle)( \hat{C}^{XY,i}_{L'} - \langle \hat{C}^{XY}_{L'}\rangle).
\end{equation}
We also show, for comparison, the theoretical error bars obtained from Equation~(\ref{eqn:delta_kg}), modified to take into account the binning. They are in generally good agreement with the MC error estimates, which, however, are slightly larger (by up to $\sim 25 \%$).


\section{Power spectra}
\label{sec:power_spectra}

\subsection{CMB Convergence-Galaxy Cross-correlation}
\label{subsec:kg_results}

\begin{figure}
\plotone{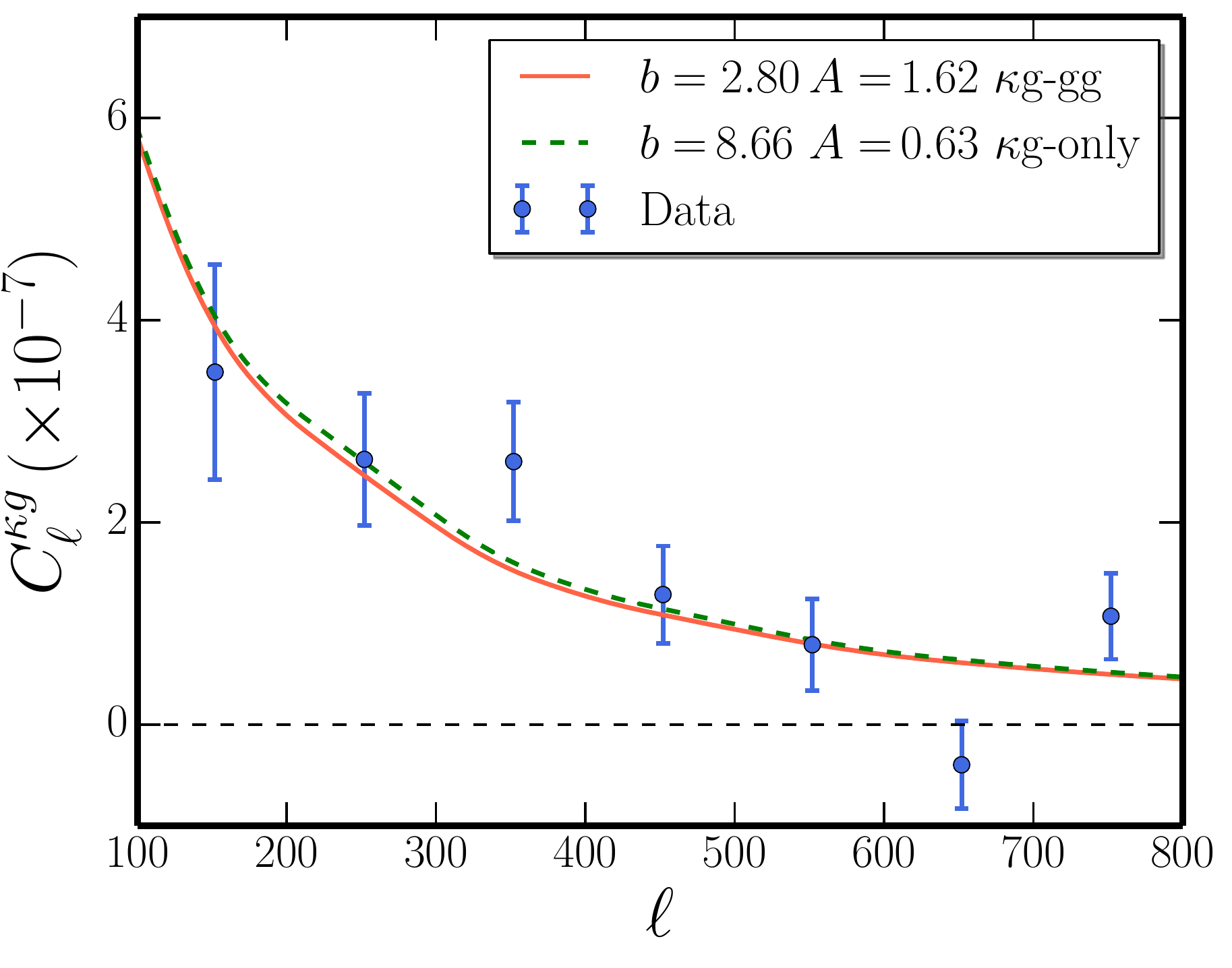}
\epsscale{0.5}
\caption{The CMB convergence-galaxy density cross-spectrum as measured from Planck and \textit{Herschel} data. The data points are shown in blue, with error bars computed using the full covariance matrix obtained from Monte Carlo realizations of convergence maps. The theoretical spectra calculated with the bias values inferred from the likelihood analysis (as described in text) using the cross-correlation data only (solid red line) and the cross-correlation together with the galaxy autocorrelation data (dot-dashed green line) are also shown; we fix $\alpha=3$ in this analysis. The null (no correlation) hypothesis is rejected at the $20\,\sigma$ level.  \label{fig:kg_all_mag}}
\end{figure}

\begin{figure} 
\plotone{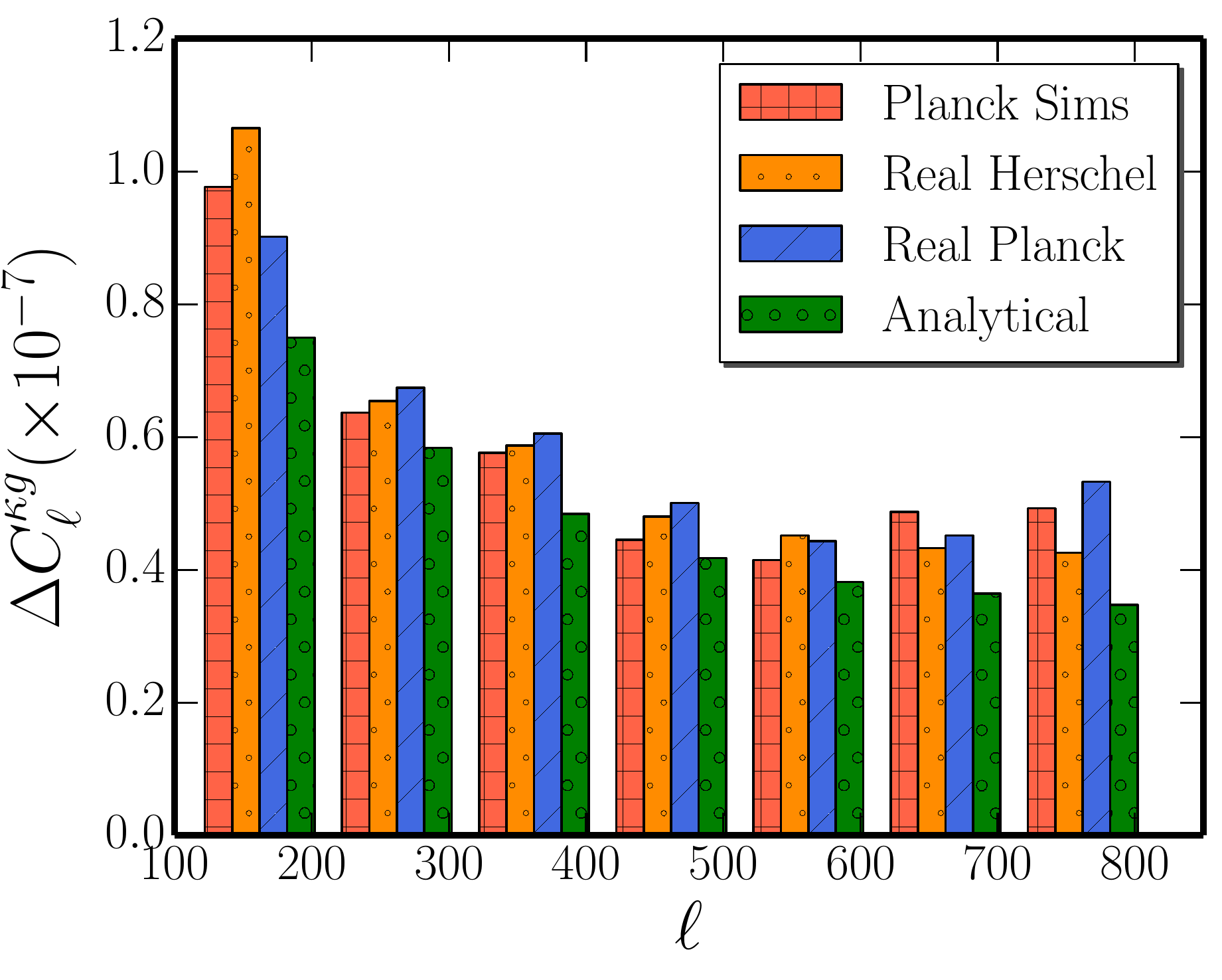}
\caption{Error estimates for the cross-power spectrum band powers. The Monte Carlo estimates associated with estimated band powers are shown in orange ($500$ simulated lensing maps correlated with the real galaxy field).  Blue bars represent errors obtained by correlating $500$ simulated galaxy maps  with the real convergence field, and the green bars represent the analytical approximation to these errors. Error estimates obtained by correlating the real galaxy field with the 100 lensing simulated maps by the Planck collaboration are shown in red. \label{fig:kg_errors}}
\end{figure}

The recovered cross-spectrum is shown in Figure~\ref{fig:kg_all_mag}.  To compute it we have applied to both maps masks that select the five H-ATLAS patches of interest. The error bars are estimated by cross-correlating 500 MC realizations of simulated CMB convergence maps (consisting of both signal and noise) with the true H-ATLAS galaxy density map, as described in Section \ref{subsec:null_tests}. This method assumes that the two maps are uncorrelated; our error estimates are a good approximation because both maps are very noisy and $C_{\ell}^{\kappa\kappa,\rm tot}C_{\ell}^{gg,\rm tot}\gg (C_{\ell}^{\kappa g})^2$. We have also estimated the errors from cross-correlations of 500 MC realizations of simulated H-ATLAS galaxy density maps with the real \textit{Planck} CMB convergence map. The former approach yields slightly smaller error bars, yet slightly larger than those estimated analytically (see Figure \ref{fig:kg_errors}). These error estimates were checked by cross-correlating  the publicly available set of 100 simulated lensing maps, which accurately reflect the \emph{Planck} noise properties, with the real H-ATLAS map. The derived error bars are comparable with those found with our baseline approach, and there is no sign of systematic under- or overestimation.

\begin{figure} 
\plotone{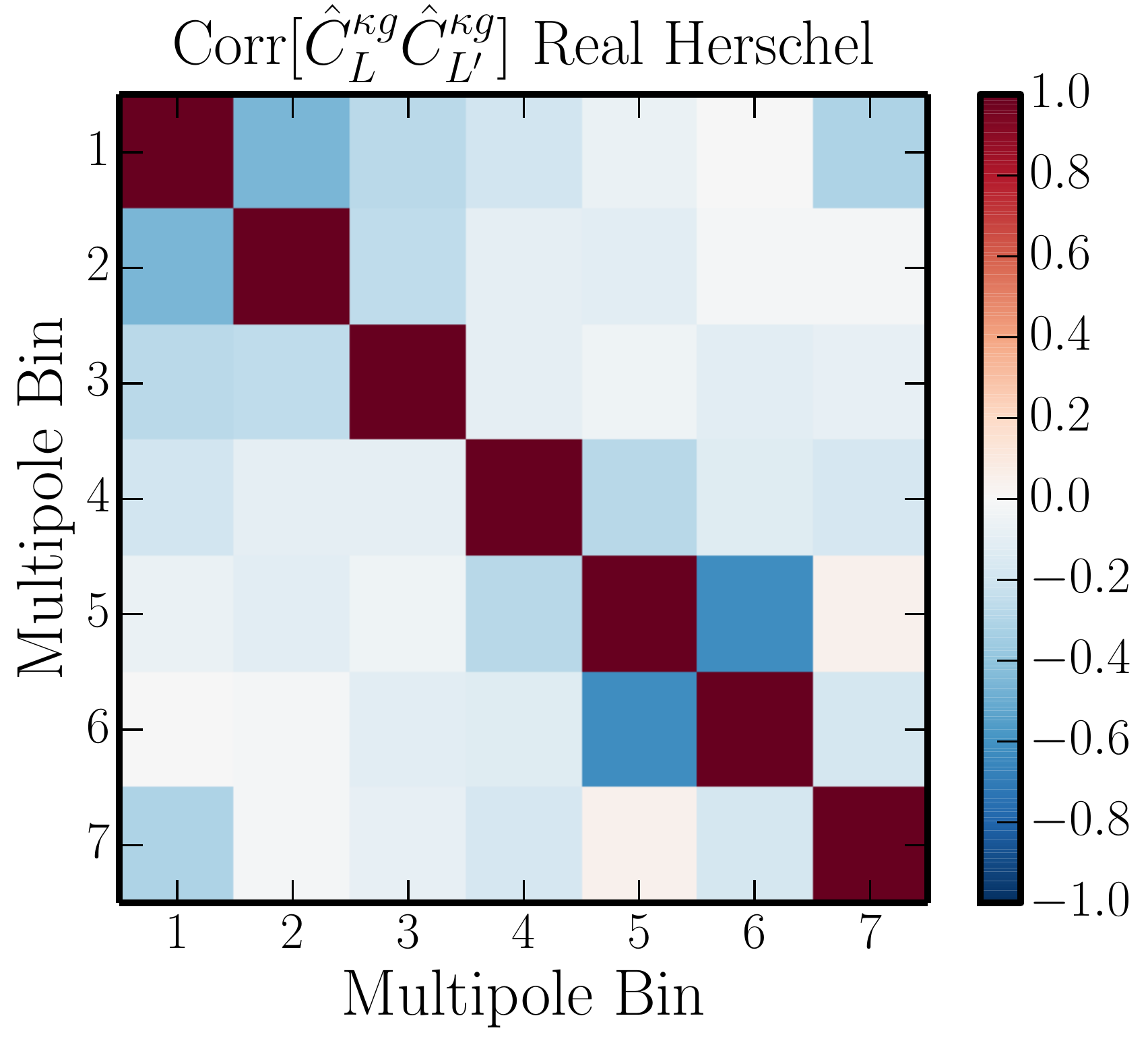}
\caption{Correlation matrix Corr$[\hat{C}^{\kappa g}_{L} \hat{C}^{\kappa g}_{L'} ]$ built from the covariance matrix obtained by correlating $500$ simulated lensing maps with the real H-ATLAS galaxy map. \label{fig:corr_kg}}
\end{figure}

We have exploited the simulations to build the covariance matrix, used to evaluate the probability that the measured signal is consistent with no correlation (our null hypothesis). As can be seen in Figure~\ref{fig:corr_kg}, the covariance matrix is dominated by the diagonal components; however, off-diagonal components are nonnegligible and have to be taken into account. The $\chi^2$ was calculated as
\begin{equation}
\chi_{\text{null}}^2 = \hat{\mathbf{C}}^{\kappa g}_{L} \,(\text{Cov}^{\kappa g}_{LL'})^{-1}\, \hat{\mathbf{C}}^{\kappa g}_{L'}.
\end{equation}
For the analysis performed with the whole H-ATLAS sample we obtained $\chi_{\text{null}}^2 = 83.3$ for $\nu = 7$ degrees of freedom (dof), corresponding to a probability that the null hypothesis holds of $p=2.89 \times 10^{-15}$.  Because the $\chi^2$ distribution has mean $\nu$ and variance $2\nu$, the null hypothesis is rejected with a significance of about $(83.3-7)/(14^{1/2})\simeq 20\,\sigma$. This is the sum in quadrature of the significance of the correlation in each band power, taking into account the correlations between different bins. The results of the $\chi^2$ analysis for each patch are reported in Table \ref{kg_sigma}.

\subsection{Galaxy Autocorrelation}\label{sec:autocorr}
\begin{figure} 
\plotone{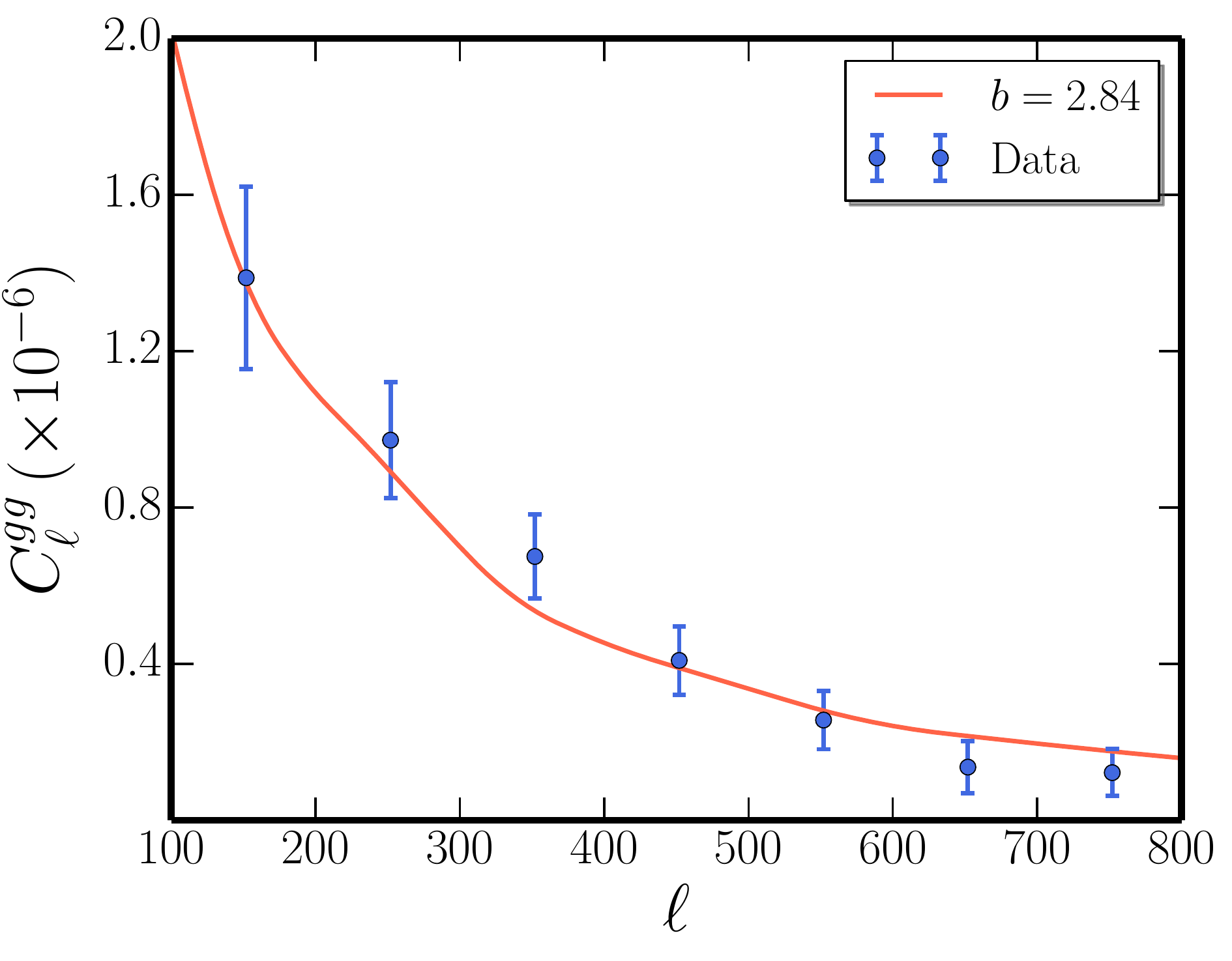}
\caption{Galaxy density autopower spectrum for the whole sample of H-ATLAS galaxies. The data points are shown in blue, and the solid (red) line is the theoretical $C_{\ell}^{gg}$ evaluated for the best-fit value of the bias obtained using a likelihood analysis on the galaxy autospectrum data.\label{fig:gg_data_all}}
\end{figure}

We also performed an analysis of the autocorrelation of \textit{Herschel} galaxies on the different patches. The shot noise subtracted autopower spectrum measured for the complete H-ATLAS data set is shown in Figure \ref{fig:gg_data_all}. The error bars on the data points are evaluated from the diagonal part of the covariance matrix built from galaxy simulations with bias $b=3$. The detected signal is highly significant ($40\,\sigma$).

\subsection{Null Tests}
\label{subsec:null_tests}
In order to verify our pipeline and the reconstructed spectra against the possibility of residual systematic errors, we performed a series of null tests, which consist of cross-correlating the real map of one field with simulated maps of the other field. Because there is no common cosmological signal, the mean correlation must be zero.

\begin{figure} 
\plotone{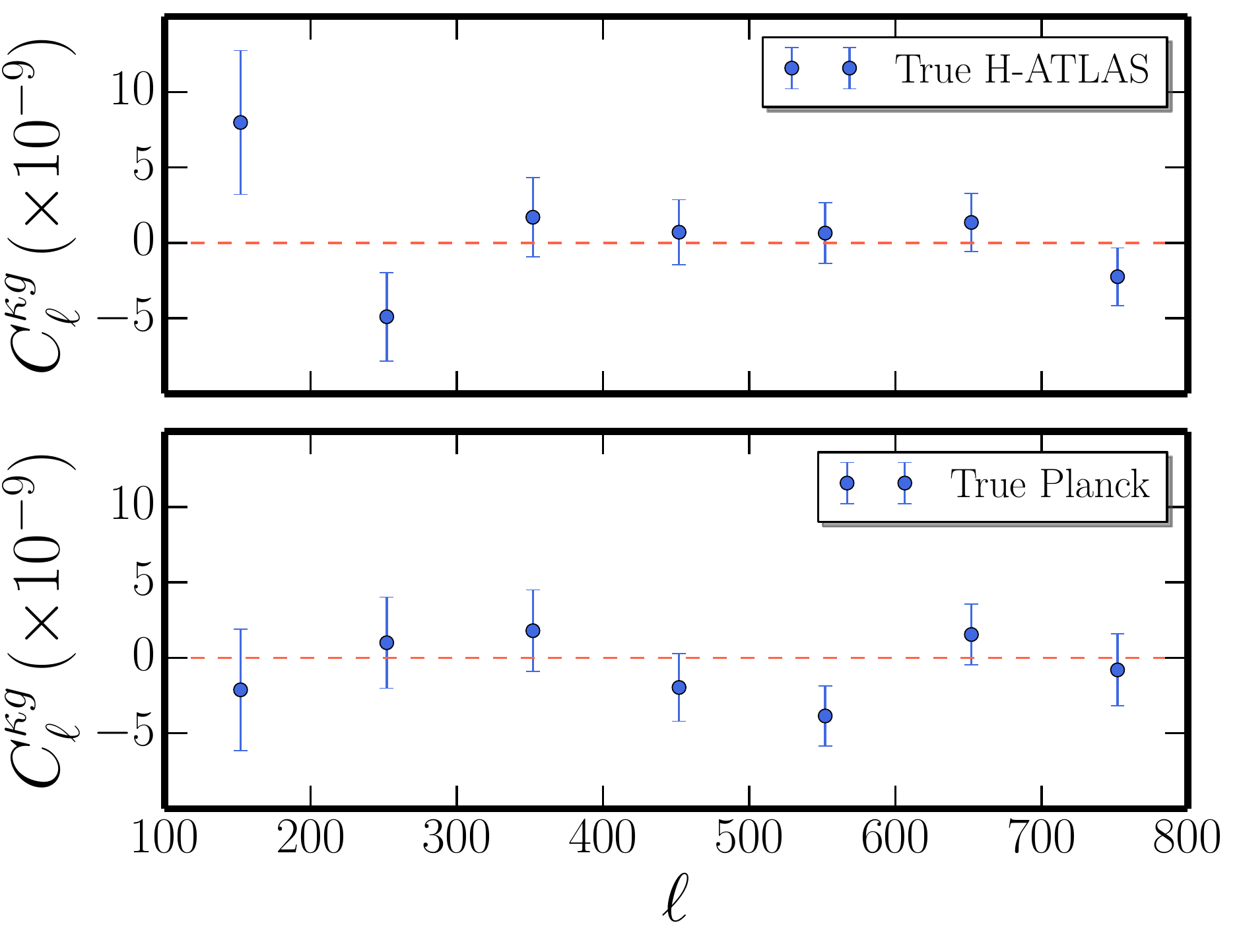}
\caption{Results of null tests. \emph{Upper panel}: mean correlation between the true H-ATLAS map including all of the five patches and $500$ simulated CMB lensing maps.
\emph{Lower panel}: mean cross-spectra between the true \emph{Planck} lensing map and $500$ simulated galaxy maps with $b=3$. No significant signal is detected in either case.
\label{fig:null_tests}}
\end{figure}

We cross-correlated our $500$ simulated CMB lensing maps (containing both signal and noise) with the real H-ATLAS galaxy density contrast map and our $500$ simulated galaxy maps constructed using $b=3$ with the true \emph{Planck}  CMB convergence map. The error bars on the cross-power spectra were computed using the covariance matrices obtained from these simulations. As illustrated in Figure \ref{fig:null_tests} in both cases no significant signal was detected. In the first test we obtained $\chi^2 = 7.2$ corresponding to a probability of the null hypothesis (no correlation) $p=0.41$, and in the second one we have $\chi^2 = 5.9$ and $p=0.55$.

A further test consisted of cross-correlating the galaxy distribution in one patch of the sky with the lensing map in another. We moved in turn the three H-ATLAS GAMA fields and the SGP field to the position of the NGP patch and shifted  the NGP galaxies to the SGP area. Then we cross-correlated each shifted galaxy map with the convergence field in the same position. The errors on the cross-correlations were obtained as above. All of the cross-spectra are consistent with no signal.

\begin{deluxetable}{cccc}
\tabletypesize{}
\tablecaption{Significance of No Cross-correlation Hypothesis Rejection\label{kg_sigma}}
\tablewidth{0pt}
\tablehead{
\colhead{Patch} & \colhead{$\chi_{\text{null}}^2 / \nu$} & \colhead{p-value} & \colhead{Significance}}
\startdata
ALL   &   $83.31/7 $&    $2.89 \times 10^{-15}  $ &    $20.3 \sigma$  \\
NGP  &   $34.03/7 $&    $1.70 \times 10^{-5}    $ &    $7.2 \sigma $ \\
SGP  &   $27.77/7 $&    $0.002                            $&    $5.6 \sigma $  \\
G09   &   $22.41/7 $&    $0.002                          $ &     $4.1 \sigma $ \\
G12   &   $22.26/7 $&    $0.002                           $ &   $4.1\sigma $  \\
G15   &   $29.23/7 $&    $1.0 \times 10^{-4}    $ &    $5.9 \sigma $ \\
\enddata
\end{deluxetable}

\section{Constraints on bias and amplitude of cross-correlation}
\label{sec:constraints}

We now discuss the cross-correlation signal of cosmological origin. Following \citet{planck_lens:2013} we introduce an additional parameter, $A$, that scales the expected amplitude of the cross-power spectrum, $C^{\kappa g}_{\ell}$, of the \emph{Planck} CMB lensing with the H-ATLAS galaxy overdensity map as $A\, C^{\kappa g}_L(b)$. Obviously, its expected value is one. Because the theoretical cross-spectrum is also basically proportional to the galaxy bias, there is a strong degeneracy between these two parameters. In order to break this degeneracy, we use also the galaxy autopower spectrum which depends only on $b$.

The best-fit values of the amplitude and of the galaxy bias were obtained using the maximum likelihood approach. In the following, we first describe the likelihood functions and present constraints on the redshift-independent galaxy bias and on the cross-correlation amplitude using galaxy autocorrelation data alone, using cross-correlation data alone, and combining both data sets. In this analysis, the cosmological parameters and the counts slope $\alpha$ are kept fixed to the fiducial values. In order to efficiently sample the parameter space, we use the Markov chain Monte Carlo (MCMC) method assuming uninformative flat priors. For this purpose we employ \textsc{EMCEE} \citep{emcee}, a public implementation of the affine invariant MCMC ensemble sampler \citep{affine_mcmc}.  In this paper, each quoted parameter estimate is the median of the appropriate posterior distribution after marginalizing over the remaining parameters with uncertainties given by the $16^{\rm th}$ and $84^{\rm th}$ percentiles (indicating the bounds of a $68\%$ credible interval). For a Gaussian distribution, as is the case when combining both data sets, these percentiles correspond approximately to $-1\sigma$ and $+1\sigma$ values, and the median of the posterior is equal to the mean and maximum likelihood value.

\begin{deluxetable*}{ccccccccccc}
\tabletypesize{}
\tablecolumns{11}
\tablewidth{0pt}
\tablecaption{H-ATLAS galaxy linear bias and cross-correlation amplitude as determined using both separately and jointly the reconstructed galaxy auto- and cross-spectra in the different patches \label{b_a_results}}
\tablehead{
\colhead{} & \colhead{$gg$} &  \colhead{} & \multicolumn{2}{c}{$\kappa g$} & \colhead{} &   \multicolumn{2}{c}{$\kappa g+gg$} &  \colhead{} &  \colhead{} &  \colhead{}  \\
\cline{2-2} \cline{4-5} \cline{7-8} \\
\colhead{Patch} & \colhead{$b$} & \colhead{} & \colhead{$b$} & \colhead{$A$} & \colhead{}  & \colhead{$b$} & \colhead{$A$} &  \colhead{} &  \colhead{$\chi^2_{\rm th}/\nu$} &  \colhead{p-value}}

\startdata
ALL & $2.84 ^{+0.12}_{-0.11}$ & \phn & $8.66^{+4.23}_{-4.37}$ & $0.63^{+0.52}_{-0.20}$ & \phn & $2.80 ^{+0.12}_{-0.11}$ & $1.62 ^{+0.16}_{-0.16}$ & \phn & $12.6/5$ & $0.03$ \\

NGP & $2.72 ^{+0.22}_{-0.21}$ & \phn & $7.92^{+5.38}_{-6.38}$ & $0.53^{+1.35}_{-0.26}$ & \phn & $2.75 ^{+0.22}_{-0.21}$ & $1.27 ^{+0.28}_{-0.29}$ & \phn & $23.1/5$ & $3 \times 10^{-4}$ \\

SGP & $2.67 ^{+0.19}_{-0.19}$ & \phn & $0.78^{+1.86}_{-0.61}$ & $3.48^{+2.63}_{-1.95}$ & \phn & $2.69 ^{+0.18}_{-0.18}$ & $1.56 ^{+0.23}_{-0.23}$ & \phn & $5.7/5$ & $0.34$ \\

G09 & $3.79 ^{+0.35}_{-0.37}$ & \phn & $8.99^{+4.02}_{-5.06}$ & $1.11^{+0.96}_{-0.36}$ & \phn & $3.72 ^{+0.35}_{-0.32}$ & $2.11 ^{+0.41}_{-0.41}$ & \phn & $6.9/5$ & $0.22$ \\

G12 & $3.43 ^{+0.35}_{-0.33}$ & \phn & $3.34^{+6.84}_{-2.55}$ & $2.04^{+3.41}_{-1.23}$ & \phn & $3.36 ^{+0.35}_{-0.33}$ & $2.05 ^{+0.47}_{-0.46}$ & \phn & $ 13.7/5$ & $0.02$ \\

G15 & $3.14 ^{+0.33}_{-0.35}$ & \phn & $8.57^{+4.85}_{-6.54}$ & $0.97^{+1.72}_{-0.38}$ & \phn & $3.13 ^{+0.34}_{-0.34}$ & $2.06^{+0.45}_{-0.47}$ & \phn & $18.4/5$ & $2\times 10^{-3}$ \\

\enddata
\end{deluxetable*}

We assumed Gaussian likelihood functions for the cross- and autopower spectra. For the galaxy autopower spectrum it takes the form

\begin{equation}
\label{eqn:like_gg}
\begin{split}
\mathcal{L}&(\hat{C}^{gg}_{L}|b) = \frac{1}{\sqrt{(2\pi)^{N_{L}}\det(\text{Cov}^{gg}_{LL'})}} \times \\
&\times\exp \Biggl\{ -\frac{1}{2} [\hat{C}^{gg}_{L} - C^{gg}_{L}(b)] \,(\text{Cov}^{gg}_{LL'})^{-1} \, [\hat{C}^{gg}_{L'} - C^{gg}_{L'}(b)]  \Biggr\},
\end{split}
\end{equation}
where $N_{L} = 7$ is the number of multipole bins and $\rm Cov^{gg}_{LL'}$ is the covariance matrix computed as described in Section~\ref{sec:cov}.

Sampling this likelihood for the measured H-ATLAS galaxy power spectrum $\hat{C}^{gg}_{L}$ we obtained constraints on the galaxy bias. Estimated values of the bias for all patches as well as for each of them  are presented in Table~\ref{b_a_results}. The results for the different patches are consistent with each other within $\lesssim 2\sigma$. The global value, $b=2.84\pm 0.12$, is consistent with earlier estimates. For example, \citet{Xia2012} found an effective value of the bias factor $b_{\rm eff}\simeq 3$ (no error given) "for the bulk of galaxies at $z\simeq 2$''. The \citet{PlanckCollaborationXXX2014} found, from their analysis of the CIB, a slightly lower value ($b_{\rm eff}\simeq 2.6$), as expected because a large contribution to the CIB comes from fainter, presumably less biased, sources.

We used the measured cross-spectra to constrain the $b$ and $A$ parameters in the same fashion. As noted above, the cross-spectra basically measure the product $A\times b$. The likelihood function is given by
\begin{equation}
\label{eqn:like_kg}
\begin{split}
&\mathcal{L}(\hat{C}^{\kappa g}_{L}|b,A) = \frac{1}{\sqrt{(2\pi)^{N_{L}}\det(\text{Cov}^{\kappa g}_{LL'})}} \times\\
&\times\exp \Biggl\{ -\frac{1}{2} [\hat{C}^{\kappa g}_{L} - A\,C^{\kappa g}_{L}(b)] \,(\text{Cov}_{LL'}^{\kappa g})^{-1} \, [\hat{C}^{\kappa g}_{L'} - A\,C^{\kappa g}_{L'}(b)]  \Biggr\},
\end{split}
\end{equation}
where $\rm Cov^{\kappa g}_{LL'}$ is the covariance matrix (Equation~(\ref{eqn:analytic_covariance})). The results are shown in Table \ref{b_a_results}.

Finally, we studied the constraints on $b$ and $A$ by combining the cross- and galaxy autospectra. For the joint analysis we used the Gaussian likelihood function that takes into account correlations between the cross- and the autopower spectra in the covariance matrix. We organized the extracted cross- and autoband powers into a single data vector as
\begin{equation}
\mathbf{\hat{C}}_{L} = (\mathbf{\hat{C}}^{\kappa g}_{L}, \mathbf{\hat{C}}^{gg}_{L}),
\end{equation}
which has 14 elements. The total covariance matrix is then written as the composition of four $7\times7$ submatrices
\begin{equation}
\text{Cov}_{LL'} =
\begin{bmatrix}
\text{Cov}^{\kappa g}_{LL'} & (\text{Cov}^{\kappa g-gg}_{LL'})^\intercal \\
 \text{Cov}^{\kappa g-gg}_{LL'} & \text{Cov}^{gg}_{LL'}  \\
\end{bmatrix}
\end{equation}
where the mixed covariance that takes into account the correlation between the two observables is
\begin{equation}
 \text{Cov}^{\kappa g-gg}_{LL'} = M_{LL_1}^{-1}P_{L_1\ell}\widetilde{\text{Cov}}^{\kappa g-gg}_{\ell\ell'}Q_{\ell'L_2}(M_{L'L_2}^{-1})^\intercal
\end{equation}
\begin{equation}
\begin{split}
&{\widetilde{\text{Cov}}^{\kappa g-gg}_{\ell\ell'}} = \frac{2}{2\ell'+1} \times \\
\times&M_{\ell\ell'}\bigl[(C^{gg}_{\ell}(b)+N^{gg}_{\ell})(C^{gg}_{\ell'}(b)+N^{gg}_{\ell'})C^{\kappa g}_{\ell}(b)C^{\kappa g}_{\ell'}(b) \bigr]^{1/2}
\end{split}
\end{equation}
In the above expressions, $\text{Cov}^{\kappa g}_{LL'}$ and $\text{Cov}^{gg}_{LL'}$ are the covariance matrices evaluated using Equation~(\ref{eqn:analytic_covariance}).

\begin{figure} 
\plotone{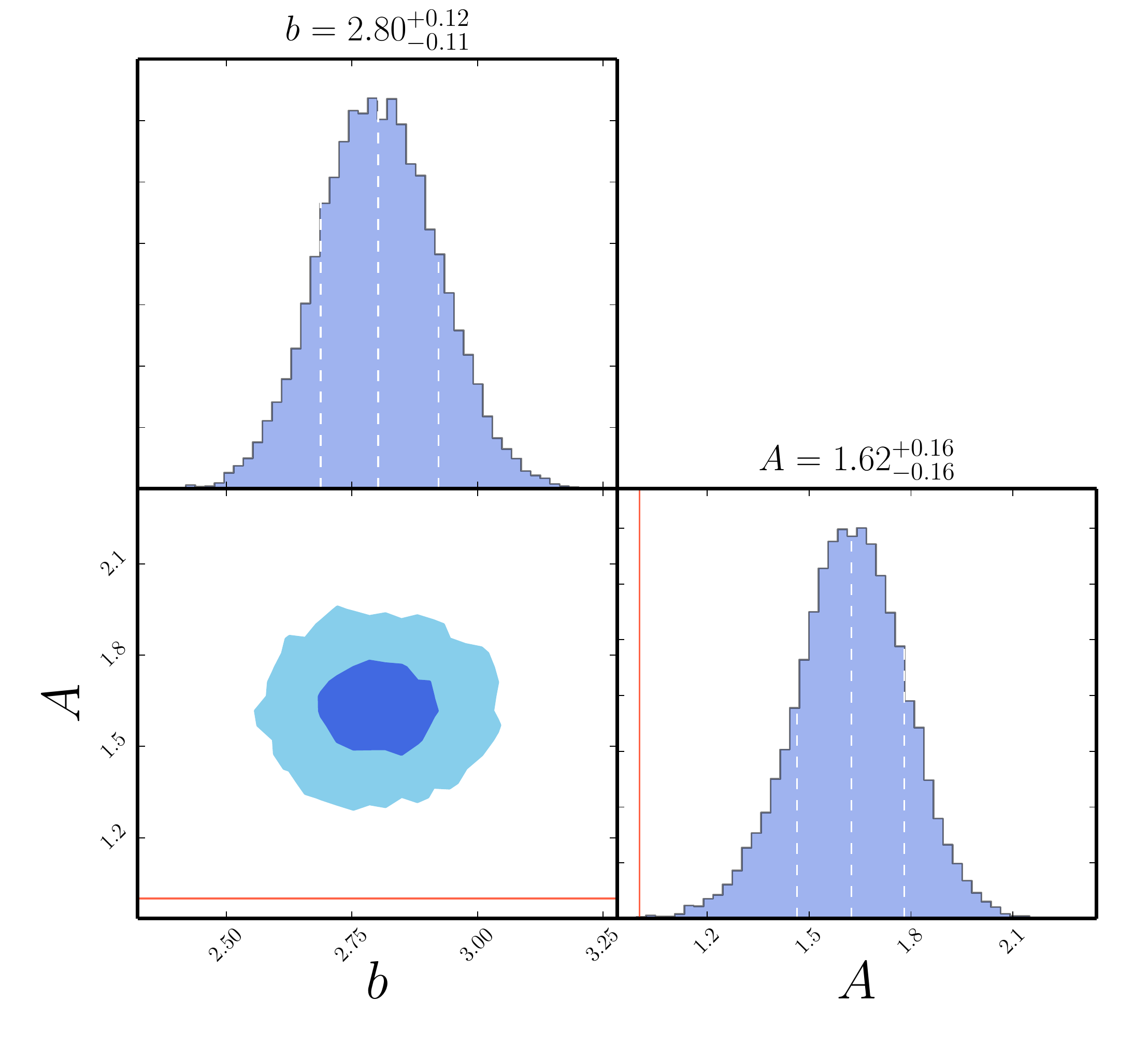}
\caption{Posterior distribution in the $b-A$ plane with the 68\% and 95\% confidence contours (darker and lighter colors, respectively), together with the marginalized distributions of each parameter with $1\sigma$ errors shown by the dashed white lines, obtained by combining the convergence-galaxy cross-correlation and the galaxy autocorrelation data for each patch. The solid red line represents the standard case in which $A=1$, and $\alpha$ is set to 3 for the analysis. \label{fig:kg_gg_pdf}}
\end{figure}

The full 2-dimensional posterior distributions of the $b$ and $A$ parameters, as well as the marginalized ones obtained from this analysis, are shown in Figure~\ref{fig:kg_gg_pdf}. Numerical values of the parameters are presented in Table \ref{b_a_results}, where the best-fit values and the errors are evaluated as the $50^{\rm th}$, $16^{\rm th}$, and $84^{\rm th}$ percentiles, respectively, of the posterior distributions. The $\chi^2$ values are evaluated as $\chi^2_{\rm th} = [\hat{\mathbf{C}}^{\kappa g}_L - A_{\rm bf}\mathbf{C}^{\kappa g}_L(b_{\rm bf})] (\text{Cov}^{\kappa g}_{LL'})^{-1}$ $[\hat{\mathbf{C}}^{\kappa g}_{L'} - A_{\rm bf}\mathbf{C}^{\kappa g}_{L'}(b_{\rm bf})] $, where $b_{\rm bf}$ and $A_{\rm bf}$ are the best-fit values. Note that the posterior distributions of $b$ and $A$ obtained using only cross-correlation data are far from being Gaussian. As a sanity check, we derived a theoretical upper limit on $A$ considering that cross-spectrum cannot be larger than the geometric mean of the two autospectra: $A \le (C^{\kappa g, \rm th}_L \rm (Cov^{\kappa g}_{LL'})^{-1} \sqrt{\hat{C}^{\kappa\kappa}_{L'}\hat{C}^{gg}_{L'}})/(C^{\kappa g, \rm th}_L \rm (Cov^{\kappa g}_{LL'})^{-1} C^{\kappa g, \rm th}_{L'}) \sim 2.5$. 

The $\chi^2$ value of the best-fit theoretical spectrum is $\chi^2_{\rm th} = 12.6$ for $\nu=5$ dof  ($\chi^2_{\rm th}/\nu = 2.5$). The significance of the detection of the theoretically expected cross-correlation signal was  evaluated as the ratio between the estimated amplitude $A$ and its error $\sigma_A$: $A/\sigma_A \simeq 10$, corresponding to a $10\sigma$ significance.

The constraint on the bias factor from the joint fit of the galaxy autocorrelation and of the cross-correlation power spectra, $b=2.80^{+0.12}_{-0.11}$, is consistent with earlier estimates \citep{Xia2012}. On the other hand, the cross-correlation amplitude is $A=1.62\pm 0.16$ times larger than expected for the standard $\Lambda$CDM model for the evolution of large-scale structure. This is at odds with the results of the cross-correlation analyses presented in the \citet{planck_lens:2013} paper, which are consistent with $A=1$ except, perhaps, in the case of the MaxBCG cluster catalog. Possible causes of the large value of $A$ are discussed in the following section.

\section{Discussion} \label{sec:discussion}

The correlation between the CMB lensing potential and the distribution of high-$z$, submillimeter selected galaxies was found to be stronger than expected for the standard cosmological model. We now address on one side the possibility that the tension between the estimated and the expected value of the amplitude $A$ is overrated because of an underestimate of  the errors and, on the other side, astrophysical effects that may enhance the measured signal.

\subsection{Noise Levels}

\begin{figure} 
\plotone{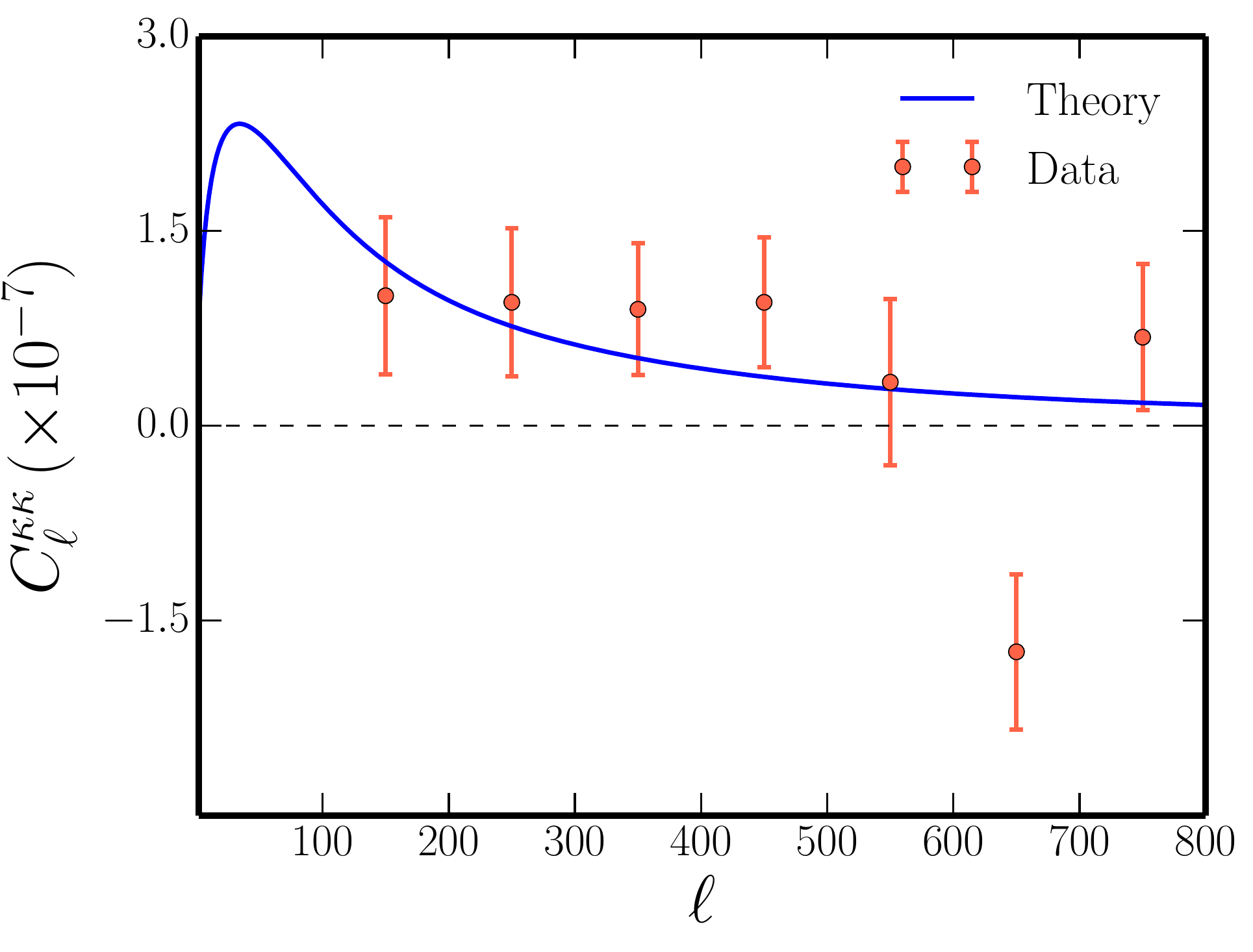}
\caption{CMB convergence autopower spectrum recovered using the H-ATLAS mask. Theory line as in Figure~\ref{fig:kk_data_planck}. \label{fig:kk_data_patches}}
\end{figure}

Due to the inhomogeneity of the noise level in the \emph{Planck} survey, the H-ATLAS patches used for the cross-correlation may have slightly higher than average effective noise. To check this possibility, we reconstructed the CMB convergence autopower spectrum for each of the H-ATLAS patches. Error bars were derived from 100 simulated \emph{Planck} lensing maps.  The results of the analysis performed combining the five patches show some excess power for $\ell \sim 400$--500  (Figure~\ref{fig:kk_data_patches}). Considering the patches separately we find that the main features of the CMB lensing power spectrum are recovered in the two largest patches, whereas the power spectrum in the three GAMA fields seems to be dominated by noise. Thus, there is an indication of a slight underestimate of the noise bias in the latter fields, but the effect on the combined patches is marginal.

\begin{figure} 
\plotone{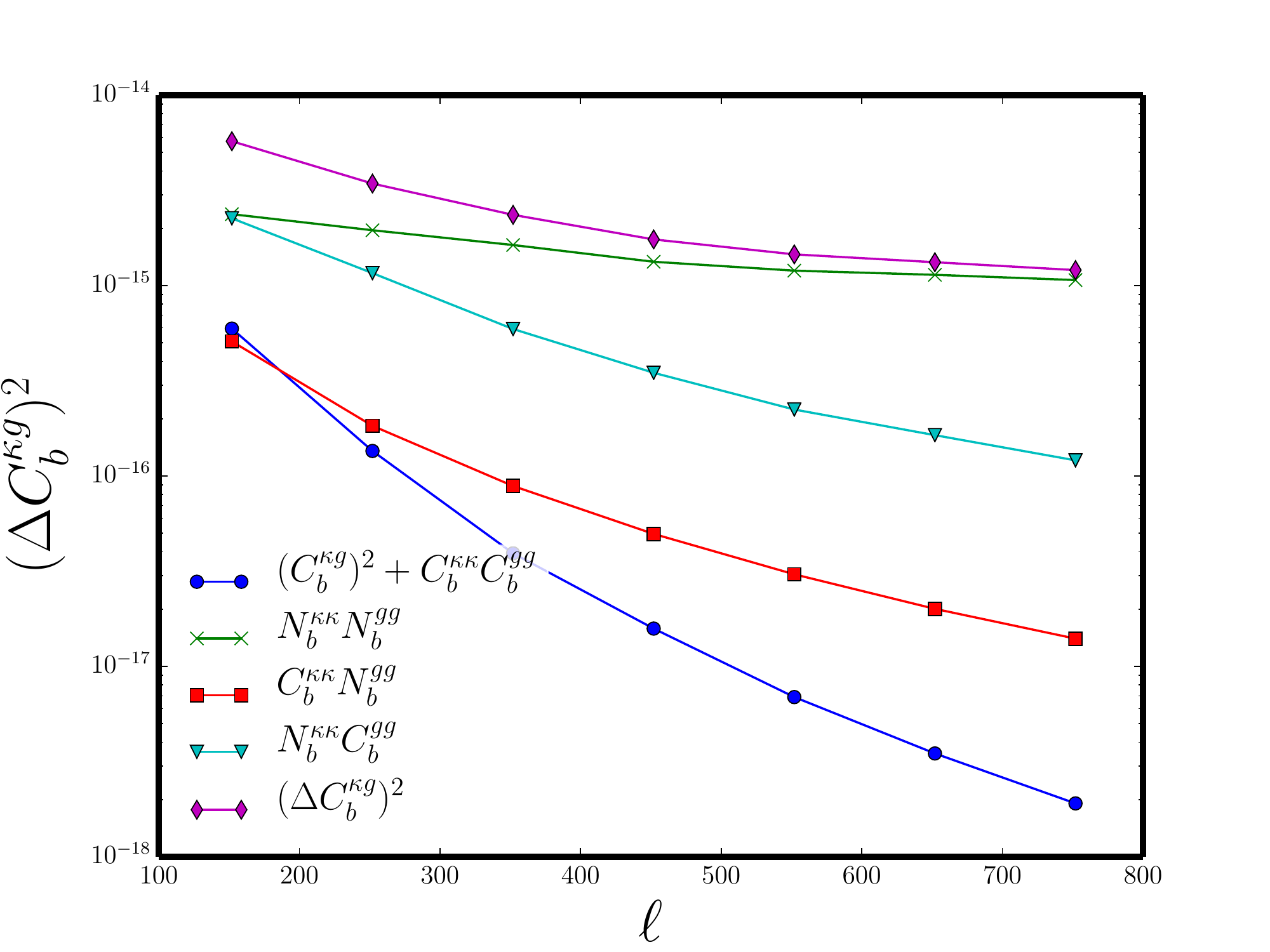}
\caption{Contributions to the cross-spectrum variance $(\Delta {C^{\kappa g}_{\ell}})^2$ [see Equation~(\ref{eq:err_budget})]. Blue line: signal only term.  Green line: noise only term. Red and cyan lines: mixed signal and noise terms. \label{fig:noise_budget}}
\end{figure}

To understand which is the main statistical error source on the cross-power spectrum, we have analyzed the contributions to the error budget. The autospectra contain a signal and a noise term as $\hat{C}^{XX}_{L} = C^{XX}_{L}+N^{XX}_L$, so that the errors on the cross-spectra can be written as
\begin{equation}\label{eq:err_budget}
\begin{split}
f_{\rm sky}(2L+1)\Delta\ell \bigl(\Delta\hat{C}^{\kappa g}_L \bigr)^2 &= \bigl[C_{L}^{\kappa\kappa} C_{L}^{gg} + (C_{L}^{\kappa g})^2 \bigr]\\
&+ N^{\kappa\kappa}_LN^{gg}_L + C_{L}^{\kappa\kappa}N^{gg}_L + C_{L}^{gg}N^{\kappa\kappa}_L.
\end{split}
\end{equation}
The first term represents the cosmic variance, the second one the pure noise, and the remaining are mixed signal-noise terms. As can be seen from Figure~\ref{fig:noise_budget}, the main contribution to the $C^{\kappa g}_L$ variance is given by the noise-only term. Moreover, the relative amplitude of the mixed terms is telling us that most of the error comes from the lensing noise. In order to reduce the errors of the reconstructed cross-spectrum, it is important to reach high sensitivity in reconstructing the CMB lensing potential. This, of course, does not include the possible systematic errors discussed below.

\subsection{Astrophysical systematics}

First we have checked the effect on the auto- and cross-spectra of errors of photometric redshift estimates. To this end we have redone the full analysis using the initial redshift distribution, $dN/dz$, i.e. the one represented by the dashed red line in Figure~\ref{fig:dndz}. We get a slightly higher value of the cross-spectrum amplitude ($A=1.70^{+0.16}_{-0.17}$) and a somewhat lower value of the galaxy bias ($b=2.59^{+0.11}_{-0.11}$). The reason for that is easily understood. As shown by Figure~\ref{fig:dndz}, the convolution of the initial $dN/dz$ with the smoothing kernel (representative of the uncertainties on estimated redshifts) results in a broadening of the distribution. This translates into a decrease of the expected amplitude for both the cross- and the autopower spectra. Hence, in order to fit the same data, we need a higher value of the galaxy bias and, consequently,  a lower value of the cross-spectrum amplitude $A$. Because the derived value of $b$ is quite sensitive to the adopted redshift distribution, the agreement with other, independent determinations implies that our $dN/dz$ cannot be badly off. Therefore, it looks unlikely that the higher than expected value of $A$ can be ascribed to a wrong estimate of $dN/dz$.

\begin{figure*} 
\plotone{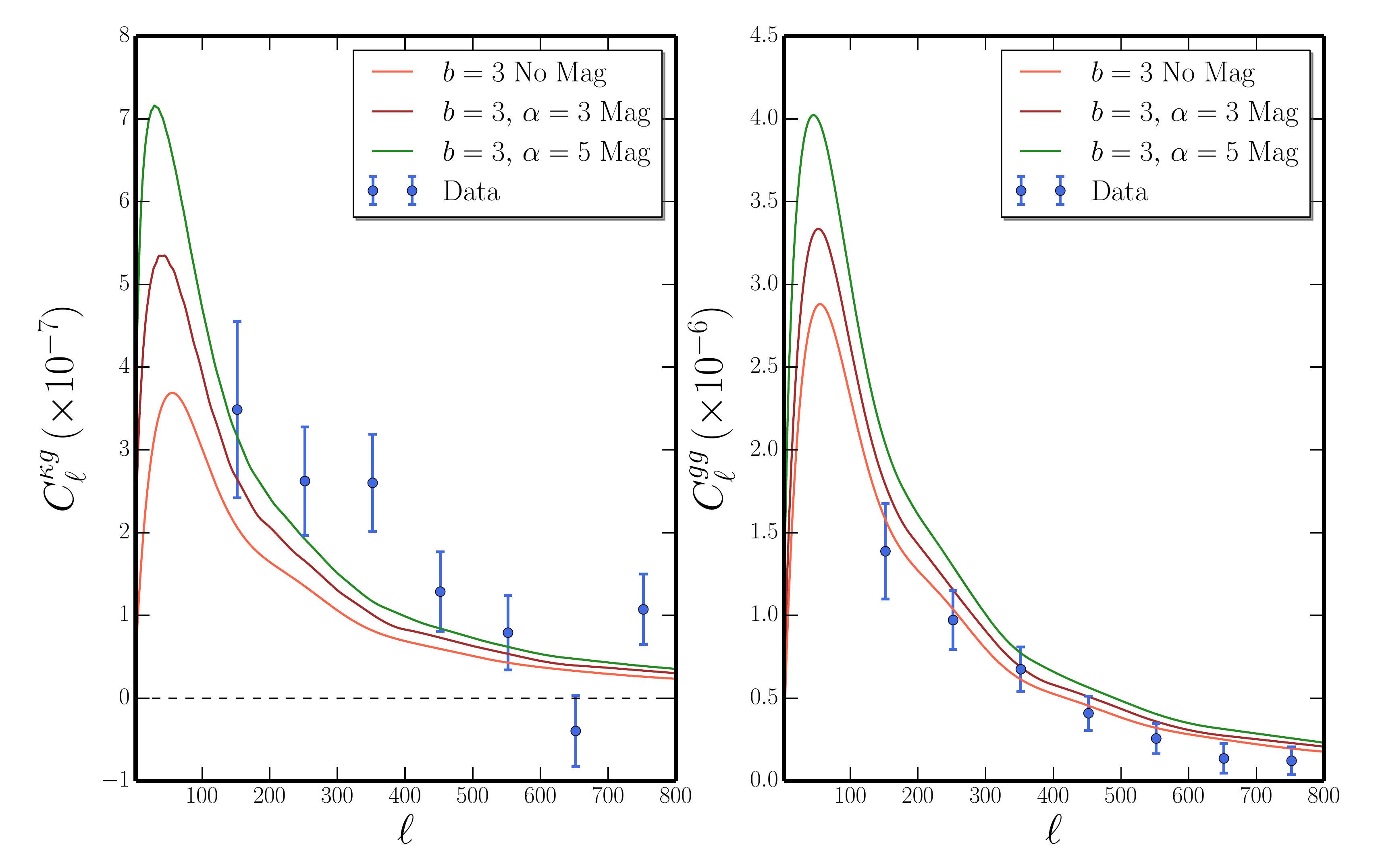}
\caption{Effect of lensing magnification bias on the cross-power spectrum (left panel) and on the galaxy autopower spectrum (right panel).  In both panels, theory lines are plotted for bias values $b=3$, while the slope of the galaxy number counts as function of flux is set to $\alpha=1$ (no magnification) and $\alpha=3,5$ as described in the legend. \label{fig:kg_gg_mag_bias}}
\end{figure*}

Our choice of a constant $b$ over the redshift range spanned by the H-ATLAS catalog is obviously an approximation, and the effective values of $b$ may be different for the cross- and the galaxy autopower spectra. To check the effect of this approximation on the estimates of $C_{\ell}^{\kappa g}$ and $C_{\ell}^{gg}$ we have computed the effective values of the bias for the two cases
\begin{equation}
\label{eqn:b_eff}
\begin{split}
b_{\rm eff}^{\kappa g} &= \frac{\int \,\frac{dz}{c}b(z)\frac{H(z)}{\chi^2(z)}W^{\kappa}(z)\frac{dN}{dz}P(k,z)}{\int\frac{dz}{c}\frac{H(z)}{\chi^2(z)} W^{\kappa}(z)\frac{dN}{dz}P(k,z)},\\
(b_{\rm eff}^{gg})^2 &= \frac{\int \,\frac{dz}{c}b^2(z)\frac{H(z)}{\chi^2(z)}(\frac{dN}{dz})^2P(k,z)}{\int\frac{dz}{c}\frac{H(z)}{\chi^2(z)}(\frac{dN}{dz})^2 P(k,z)},
\end{split}
\end{equation}
using the bias evolution model $b(z)$ from \cite{sheth:1999} for halo masses in the range $10^{12}\hbox{--} 10^{13}\,\hbox{M}_{\odot}$. We find that $b_{\rm eff}^{\kappa g}$ is only slightly larger (by $\simeq 6\%$) than $b_{\rm eff}^{gg}$. Hence, considering a redshift-dependent bias factor would only marginally affect the expected cross-spectrum.

Weak lensing by foreground structures modifies the \textit{observed} density of background sources compared to the real one \citep[magnification bias;][]{ho08,xia09} and is especially important for high-redshift objects. The effect on the galaxy overdensity kernel is described by the second term on the right-hand side of Equation~(\ref{eqn:wg}). The effect of the magnification bias on both $C_{\ell}^{\kappa g}$ and $C_{\ell}^{gg}$ is illustrated in  Figure~\ref{fig:kg_gg_mag_bias} where we show the expected power spectra for $A=1$, $b=3$, and three values of $\alpha$: 1 (no magnification bias), 3, and 5. The impact of the magnification bias is clearly stronger for $C_{\ell}^{\kappa g}$.

\begin{figure} 
\plotone{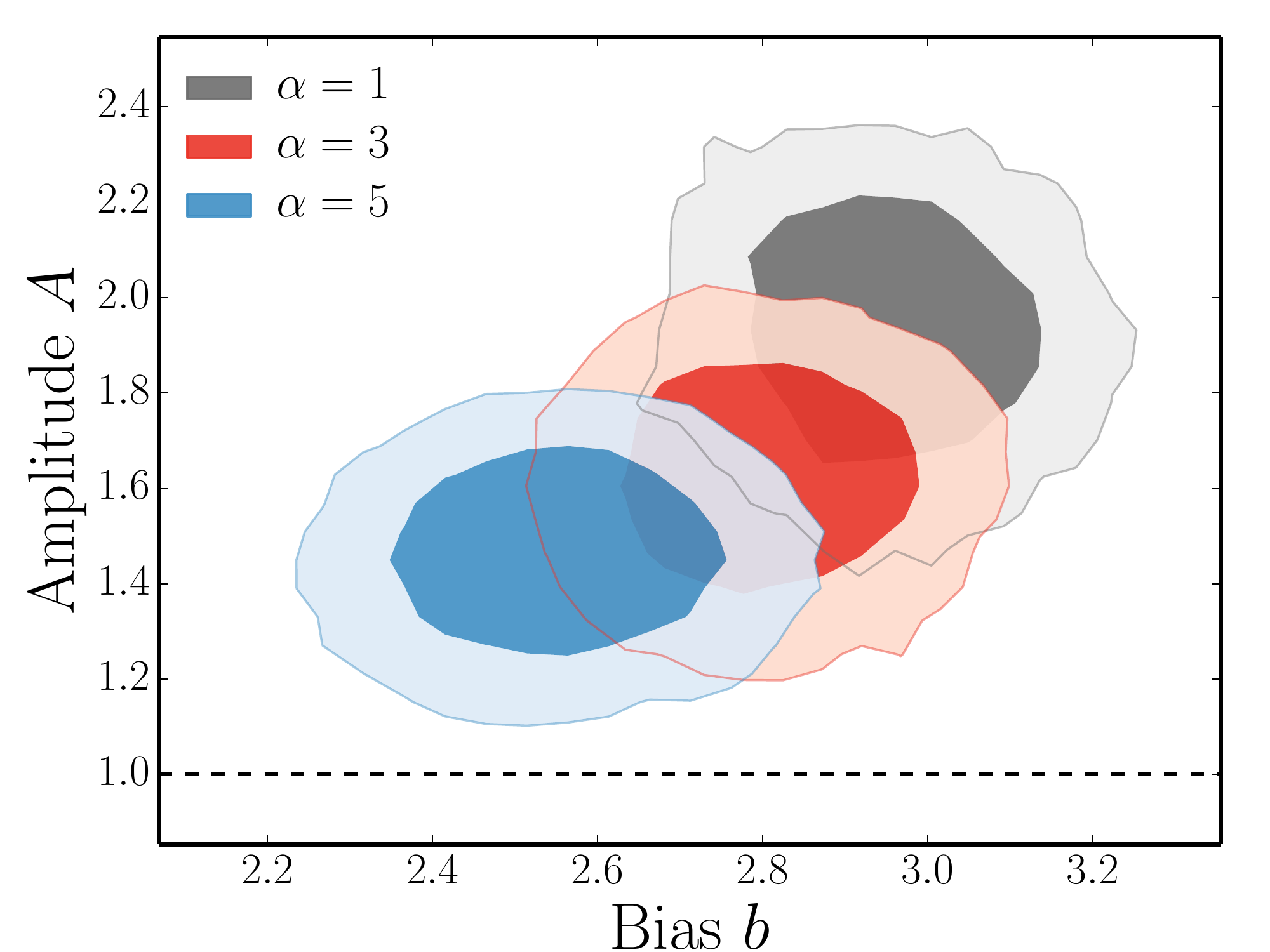}
\caption{Effect of fixed slope of number counts $\alpha$ on the inferred values of cross-correlation amplitude $A$ and bias $b$. We show $1-$ and $2\sigma$ contours (darker and lighter shaded regions, respectively). As the $\alpha$ parameter increases, both $A$ and $b$ shift toward smaller values. \label{fig:kg_gg_alphas_pdf}}
\end{figure}

Fitting the joint data for $\alpha=1$ we find  $b=2.95^{+0.12}_{-0.11}$ and  $A=1.93^{+0.18}_{-0.19 }$ while for $\alpha=5$ $b=2.55^{+0.13}_{-0.12}$ and  $A=1.46 \pm 0.14$. The contour plots in the $A-b$ plane are shown in Figure~\ref{fig:kg_gg_alphas_pdf}. Higher values of $\alpha$ imply lower values of $A$, but even for $\alpha = 5$ the data require $A>1$.

Another systematic effect that can bias our measurement of the CMB convergence-galaxy cross-correlation is the leakage of cosmic infrared background (CIB) emission into the lensing map through the temperature maps used for the lensing estimation, as it correlates strongly with the CMB lensing signal \citep{planck_cib:2013}. The 857\,GHz \emph{Planck} map used by \citet{planck_lens:2013} as a Galactic dust template also removes the portion of the CIB fluctuations that have a spectral index similar to that of Galactic dust. However, as noted in that paper, this approach is liable to problems due, for example, to variation of Galactic dust spectral indices across the sky, as well as to the mismatch between the beams at 100/143/217 and 857 GHz.

The H-ATLAS galaxies are well below the \emph{Planck} detection limits \citep[their flux densities at 148\,GHz are expected to be in the range 0.1--1\,mJy, hence are much fainter than sources masked by][]{planck_lens:2013}. Thus they are part of the CIB measured by \emph{Planck}. If they are only partially removed by the use of the 857\,GHz map, they are potentially an important contaminant of the cross-correlation, resulting in an enhancement of the observed signal. The shot-noise correction applied by the \emph{Planck} team removes only partly the contamination by infrared sources because their main contribution to the fluctuation field is due to clustering.

Estimates of biases to the lensing reconstruction signal from extragalactic sources have been worked out by the \citet{osborne:2013,vanEngelen:2013}. However, a calculation of the bias on the cross-spectrum discussed in this paper is beyond the scope of the present paper.  We expect that with the next release of the \textit{Planck} data, CMB lensing maps at different frequencies will become available. This will allow us to investigate the CIB leakage issue in more detail.

Clusters of galaxies, which trace the large-scale potential responsible for the CMB lensing, are visible at millimeter and submillimeter wavelengths via the scattering of CMB photons by hot electrons {(Sunyaev-Zel'dovich effect)} and might therefore contaminate the cross-correlation signal to some extent. However, the redshift range populated by galaxy clusters only marginally overlaps with the redshift distribution of our sources, so that this contamination is negligible.

\begin{figure} 
\plotone{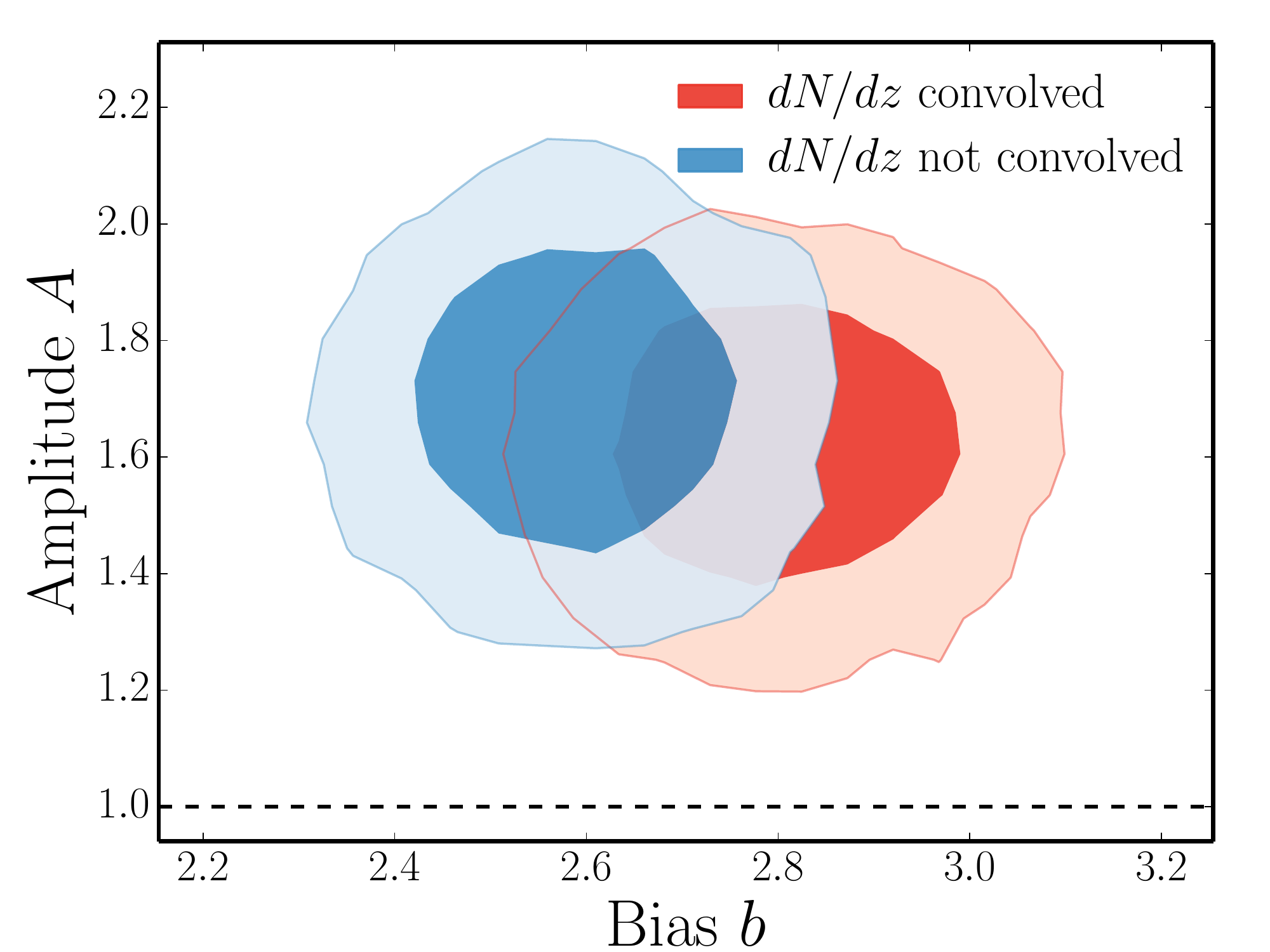}
\caption{Posterior distributions for $A$ and $b$ obtained using the convolved (red contours) and the unconvolved $dN/dz$  (blue contours).  \label{fig:kg_gg_un_convolved_pdf}}
\end{figure}

\section{Summary and conclusions}
\label{sec:conclusions}
We have presented the first measurement of the correlation between the lensing potential derived from the \emph{Planck} data and a high-$z$ ($z\ge 1.5$) galaxy catalog from the \emph{Herschel}-ATLAS survey, the highest redshift sample for which the correlation between \emph{Planck} CMB lensing and tracers of large-scale structure has been investigated so far.
We have shown that the expected signal is remarkably strong, in spite of the small area covered by the H-ATLAS survey (about 1.3\% of the sky), suggesting that cross-correlation measurements between CMB lensing maps and galaxy surveys can provide powerful constraints on the evolution of density fluctuations, on the nature of the dark energy, and on properties of tracers of the matter distribution, provided that a good control of systematic errors  for both data sets can be achieved.

The null hypothesis (no correlation) was rejected with a significance of about $20\,\sigma$ and the significance of the detection of the theoretically expected cross-correlation signal was found to be $10\,\sigma$. The reliability of this result was confirmed by several null tests. A joint analysis of the cross-spectrum and of the autospectrum of the galaxy density contrast yielded a galaxy bias parameter of $b=2.80^{+0.12}_{-0.11}$, consistent with earlier estimates for H-ATLAS galaxies at similar redshifts. On the other hand, the amplitude of the cross-correlation was found to be a factor $1.62 \pm 0.16$ higher than expected from the standard model and found by cross-correlation analyses with other tracers of the large-scale structure.

We have investigated possible reasons for the excess amplitude. Some of them, such as the redshift dependence of the bias parameter or the contamination by the Sunyaev-Zel'dovich effect, were found to be negligible. Others, such as the magnification bias due to weak gravitational lensing or errors in the photometrically estimated redshifts, can contribute significantly to the observed excess but cannot fully account for it. A possible culprit is some residual contamination of convergence maps by unresolved infrared sources \citep{osborne:2013,vanEngelen:2013}, adding a substantial contribution to the measured correlation between the lensing convergence and the H-ATLAS high-$z$ sources, which are unresolved by \emph{Planck}. However, a detailed calculation of this effect is complicated and beyond the scope of the present paper.

We have also investigated the possibility that the tension between the observed and the expected cross-correlation amplitude was overrated because the noise level of the convergence maps in the regions used for the cross correlation is above typical values. This turned out to be the case in the three GAMA fields, but the effect on the combination of fields was found to be marginal.

An exquisite mapping of the CMB lensing pattern is one of the major goals of operating and planned CMB probes because of its relevance in studying cosmological structure formation and the properties of the dark energy. Forthcoming data releases by Planck as well as future CMB lensing measurements from suborbital probes will be most relevant to further address the results presented here and improve the constraining power of these studies, both in a cosmological and astrophysical context.

\acknowledgments
We thank Karim Benabed and Laurence Perotto for useful discussions and comments, Duncan Hanson and Michal Michalowski for a careful reading of the paper, and the anonymous referee for insightful comments that helped us improve the paper. F.B. would like to thank Simone Aiola, Matteo Calabrese and Giulio Fabbian for stimulating discussions. C.B. thanks Andrew Jaffe and Radek Stompor for useful discussions. We gratefully acknowledge support from INAF PRIN 2012/2013 ''Looking into the dust-obscured phase of galaxy formation through cosmic zoom lenses in the Herschel Astrophysical Terahertz Large Area Survey'', and from ASI/INAF agreement 2014-024-R.0. F.B. acknowledges partial support from the INFN-INDARK initiative. L.D., R.J.I. and S.M. acknowledge support from the European Research Council (ERC) in the form of Advanced Investigator Program, COSMICISM. J.G.N. acknowledges financial support from the Spanish CSIC for a JAE-DOC fellowship, cofunded by the European Social Fund. The work has been supported in part by the Spanish Ministerio de Ciencia e Innovacion, AYA2012-39475-C02-01, and Consolider-Ingenio 2010, CSD2010-00064, projects.
The authors acknowledge the use of CAMB and HEALPix packages and of the \emph{Planck} Legacy Archive (PLA). A.L. thanks SISSA for warm hospitality.


\end{document}